\documentclass[10pt,a4paper,headings=normal,oneside,numbers=noendperiod,DIV=9]{scrartcl}

\usepackage[english]{babel}
\usepackage{tabularx}
\usepackage{amsmath}
\usepackage{graphicx}
\usepackage{subfigure}
\usepackage{gensymb}
\usepackage{array}
\usepackage{longtable}
\usepackage[numbers]{natbib}
\usepackage{enumitem}

\newcolumntype{f}[1]{>{\raggedright \hspace{0pt}}p{#1}}

\setkomafont{disposition}{\bfseries\rmfamily}

\title{\Large Rectification of position data of Scotland in Ptolemy's \textit{Geographike Hyphegesis}\thanks{Author-generated postprint; published in Survey Review (2014) 46: 231--244, DOI\newline 10.1179/1752270613Y.0000000085.}}

\author{\large Christian Marx\thanks{C. Marx, Gropiusstra{\ss}e 6, 13357 Berlin, Germany; e-mail: ch.marx@gmx.net.}}

\date{}

\begin{document}

\maketitle

\begin{abstract}
\noindent
\textbf{Abstract:} The ancient geographic coordinates given for places of Great Britain in Ptolemy's \textit{Geographike Hyphegesis} are investigated by means of geodetic methods.
The turning of Scotland to the east is modelled by a three-dimensional rotation.
On the basis of different data sets of control points, the parameters of the rotation are estimated by means of methods of adjustment theory.
Furthermore, a geodetic-statistical analysis method is applied to Scotland, by which groups of places of homogenous distortions and modern counterparts of the ancient places are determined.
Based on the results of the investigations, answers are given for questions concerning Ptolemaic positions unsolved so far.

\end{abstract}

\section{Introduction}

The oldest comprehensive description of Great Britain that has been handed down can be found in Book II of the \textit{Geographike Hyphegesis} (GH) by Klaudios Ptolemaios (Ptolemy, ca. 100--178).
That description of Great Britain, which he called \textit{Albion}, is part of a location catalogue (GH Books II--VII), wherein positions of about 6300 places of the whole \textit{Oi\-kou\-me\-ne} (the inhabited world known to the Greeks and Romans) are given by means of geographic coordinates in Ptolemy's geographic reference system, which differs from the modern system by its zero meridian at the 'Blest islands' (GH IV.6.34).

For different fields of research, the modern counterparts of unknown places of the GH have been of interest as well as the accuracy of the Ptolemaic coordinates and their origination.
Ptolemy's description often shows considerable differences to the actual situation. In particular, Ptolemaic Scotland, the part of Great Britain north of Hadrian's Wall, is turned to the east.
If the Ptolemaic positions are not rough, conjectural positions but locality determinations based on accurate data sources (such as military measurements), it can be expected from the Ptolemaic coordinates that they are systematically distorted.
The determination of systematic errors provides a rectification and the possibility of identifying unknown Ptolemaic places.

The first data on Great Britain possibly originate from the Greek Pytheas of Massalia, who circumnavigated it in ca. 330~B.C. and traveled to the legendary \textit{Thule}.
A major source of the GH was the works of the Greek geographer Marinos of Tyre (ca. 70--130), which are dealt with in detail in Book I of the GH.
Presumably, military sources were available to Marinos (cf. \citep{tie59}).
Geographic information arose with the Roman conquest of Great Britain in the first century.
Roman sources were surely available to Ptolemy, which is affirmed by the occurrence of Latin place names in the GH and by the high accuracy of the Ptolemaic data determined by Marx \citep{mar11b}, Kleineberg et al. \citep{kle12}, and Marx and Kleineberg \citep{markle12}.
Agricola, the Roman governor of the province \textit{Britannia}, conquered parts of Scotland.
These regions, however, were given up later; the province \textit{Britannia} was bordered by Hadrian's Wall.
Two known seafarings surely yielded geographic information on Scotland: the circumnavigation of Scotland by the geographer Demetrius in 81--83 mentioned by Plutarch (cf. \citep{jon96}) and the circumnavigation of Great Britain by Agricola's fleet in 84 mentioned by Tacitus.

Ptolemy's places of \textit{Albion} have been the subject of a multitude of investigations so far; Strang \citep{str97,str98b} gives an overview.
Tierney \citep{tie59} discusses the works of Ptolemy's predecessors with regard to the influence on Ptolemy's \textit{Albion}.
Thomas \citep{tho75} identifies the Ptolemaic places in Scotland by a comparison of the Ptolemaic distances with the true distances from place to place.
Richmond \citep{ric22} corrects the turning of Scotland by means of a rotation by 90\degree, performed by an exchange of longitude and latitude. The essential work by Rivet and Smith \citep{riv79} deals with the history and literary sources concerning ancient places in Great Britain; the turning of Scotland is explained by a rotation by ca. 50\degree.
Strang \citep{str97,str98b} gives a comprehensive analysis of the distortions of \textit{Albion} on the basis of mappings of modern and Ptolemaic positions, which is discussed in Section \ref{sec:england}.
He describes the distortions by rotations, scaling errors and shifts. However, the results of the investigations of Kleineberg et al. \citep[p. 35 ff.]{kle12} on England and the present investigations on Scotland show that the presence of more than one rotation is doubtful.

The objective of an analysis of the distortions of Ptolemy's Scotland should be to explain them realistically and as simply as possible.
In Section \ref{sec:schottland} the turning of Scotland is described by a three-dimensional (3D) rotation, which appears to be a satisfactory modeling of the turning of Scotland.
The pivot point and the rotation angle are determined by methods of adjustment theory.
Further distortions of the Ptolemaic positions are determined along with their identifications by means of a geodetic-statistical analysis method, which is described introductorily in Section \ref{sec:methode}.
In Section \ref{sec:england} results of the analysis of Ptolemaic England are given, which are of importance for the investigation of Scotland.

\section{Analysis method} \label{sec:methode}

Because of the unreliability and inaccuracy of the ancient coordinates of the GH, further information must be consulted for the identification of the ancient places in addition to a computational analysis of the coordinates, e.g. historical information, archaeological sites, and toponymy.
According to this, Ptolemy's data for Europe (GH Books II, III) have been investigated interdisciplinarily, whereby identifications of the Ptolemaic places have been affirmed and newly found and the errors and accuracy of the coordinates have been determined (see \citep{mar11b}, \citep{kle12}, \citep{markle12}).
The underlying analysis method is described in detail by Marx \citep{mar12a} and is therefore dealt with only briefly in the following.

Investigations of regions with a multitude of known Ptolemaic places (e.g. \textit{Italia} in GH III.1, see \citep[p. 10 ff.]{markle12}) revealed that the places subdivide into groups with systematic distortions; these are scaling errors and shifts.
One exception among the investigated Ptolemaic places of Europe is Ptolemaic Scotland, where additionally a rotation can be found.
This case is introduced only in Section \ref{sec:drehung}.
Furthermore, the Ptolemaic coordinates have random components, which originate from errors and inaccuracies in the data used by Ptolemy (measurement data, information from travel reports, maps) and from Ptolemy's determination of geographic coordinates based on his data sources (on the sources see \citep[p. 16 ff.]{stu06}, \citep[p. 5 ff.]{kle12}).

The aim of the geodetic-statistical analysis of the Ptolemaic coordinates is the determination of groups of places of homogenous distortion (transformation units).
The analysis method is based on adjustment theory and statistical hypothesis testing.
The observation equations of the applied Gau\ss{}-Markov model (see e.g. \citep[p. 117 ff.]{nie02}) are
\begin{equation} \label{eqn:beogl}
\begin{aligned}
\Lambda_{i} + v_{\Lambda\,i} &= m_{\lambda} \, \lambda_{i} + \Lambda_{0k} \\
\Phi_{i} + v_{\Phi\,i} &= m_{\phi} \, \phi_{i} + \Phi_{0k} \; ,
\end{aligned}
\end{equation}
where the Ptolemaic longitude $\Lambda_i$ and latitude $\Phi_i$ of a place with index $i$ are observations, the modern longitude $\lambda_i$ and latitude $\phi_i$ are constants, the scale parameters $m_{\lambda}$ and $m_{\phi}$ are unknowns or constants (see below), the shift parameters $\Lambda_{0k}$ and $\Phi_{0k}$ of a group of places with group index $k$ are unknowns, and $v_{\Lambda\,i}$ and $v_{\Phi\,i}$ are residuals (corrections) taking into account random errors.
Model (\ref{eqn:beogl}) describes a transformation of the modern into the ancient coordinates of a place.
The transformation parameters $m_{\lambda}$, $m_{\phi}$, $\Lambda_0$, $\Phi_0$ contain local and global effects.

The scale parameters $m_{\lambda}$, $m_{\phi}$ are assumed to be spaciously valid.
Scalings may originate from Ptolemy's overestimation of the longitudinal dimension of the \textit{Oikoumene} as well as from differences between ancient measurement units, which were unintentionally not considered.
Owing to interactions of different influences, $m_{\lambda}$ and $m_{\phi}$ are possibly not entirely identical (see the example of \textit{Peloponnesus} in GH III.16 given in \citep[p. 125 f.]{markle12}).
Inconsistencies of the ancient coordinates and disadvantageous geometries of groups of places of homogenous distortions can adulterate the adjustment of transformation parameters such that the results are unrealistic.
Thus, for $m_{\lambda}$ and $m_{\phi}$ approximate values are determined, which are used as constants in several steps of the analysis method and are iteratively improved.

The parameter $\Lambda_{0k}$ contains the difference between the Ptolemaic and the modern zero meridian.
The computed $\Lambda_{0k}$ and $\Phi_{0k}$ are in general no real shifts.
In order to illustrate the actual shifts, relative shifts of transformation units with respect to a chosen transformation unit are determined by means of
\begin{equation} \label{eqn:trdiff}
\begin{aligned}
\Delta \Lambda_{0k} &= \Lambda_{0k} - \Lambda_{0\mathrm{R}} \\
\Delta \Phi_{0k} &= \Phi_{0k} - \Phi_{0\mathrm{R}} \; .
\end{aligned}
\end{equation}
$\Lambda_{0\mathrm{R}}$ and $\Phi_{0\mathrm{R}}$ are the adjusted parameters of the transformation unit which is taken as a reference.

In the Greek manuscripts of the GH the coordinates are listed by way of Milesian numerals; they are given in degree and fractions of degree. The smallest resolution occurring is $\frac{1}{12}\degree = 5'$.
Marx \citep{mar11a} shows that the actual resolution is partly lower and gives a method for the estimation of the occurring resolutions.
According to the resolution, the standard deviations $\sigma_{\Lambda\,i}$ and $\sigma_{\Phi\,i}$ of the $\Lambda_i$ and $\Phi_i$ are chosen in the stochastic part of the adjustment model; of the most accurate coordinates ca. 5$'$ are assumed.
Correlations between coordinates are not considered because there is no information about dependencies.
Coordinate values not explicable by the distortion model (\ref{eqn:beogl}) are regarded to be grossly erroneous;  often they can be explained by a scribal error in the manuscripts.

In essence, the analysis method is a multi-stage combinatorial search for transformation units.
In a combinatorial search for consistent subsets of data, different combinations of the observations are tested for whether they satisfy specific conditions, e.g. a statistical test (for other applications of such a strategy see \citep{nei05}, \citep{koc07}).
According to model (\ref{eqn:beogl}), the quantities to be combined are the Ptolemaic coordinates $\Lambda_{i}$ and $\Phi_i$; however, the combinatorial search is extended to the modern positions ($\lambda_i$, $\phi_i$) because more than one (uncertain) identification can be given for a place.
Moreover, the problem is complicated by differences between ancient coordinate values in the manuscripts.
The manuscripts are presumably based on the two recensions $\Omega$ and $\Xi$; $\Xi$ is only represented by the manuscript Codex Vaticanus Graecus 191 (X).
An edition of both recensions is published by St\"uckelberger and Gra\ss{}hoff \citep{stu06}, which has been used for the investigations.
By means of the analysis method an inconsistent input-variant of a coordinate is replaced by a consistent variant, if available.

For an area under investigation, the procedure of the analysis method is the following:
\begin{itemize}[itemsep=-3ex,leftmargin=0.5cm]
\item[1.] Initial solution:
    analysis of the resolution of the ancient coordinate values,
    determination of approximate values for the transformation parameters, generation of initial
    subsets of places with similar distortions by means of a visualisation of residuals (cf. Fig. \ref{fig:strang}).\\
\item[2.] Combinatorial search:
    search for transformation units in the initial subsets; multiple identifications per place are possible; statistical tests: overall model test of the adjustment model, individual test for gross errors in the coordinates.\\
\item[3.] Forward-strategy:
    search for the best possible mergings of unassigned places with nearby transformation units; multiple identifications and ancient coordinate variants per place are possible; statistical tests: individual test, final overall model test; geometric tests: distance test, point-in-polygon test concerning the convex hull of a transformation unit.\\
\item[4.] Verification of the scales:
    test of the suppositional scales introduced in step 1 for validity by an adjustment; statistical test: t-test.\\
\item[5.] Merging of transformation units:
    combinatorial search for possible mergings of neighbouring transformation units; statistical test: analysis of variance; geometric tests: distance, overlap by means of point-in-polygon test.\\
\item[6.] Postprocessing:
    if present, test of topographically implausible assignments of places to transformation units for possible rearrangements; statistical tests: overall model test, individual test.\\
\end{itemize}
(On the applied tests see e.g. \citep[pp. 66, 150, 171 ff., 356]{nie02}, \citep[pp. 189, 193]{jae05}, \citep[p. 28]{bil96}.)

Based on the determined transformation units, presumable modern coordinates $\bar{\lambda}_i$, $\bar{\phi}_i$ can be computed for unidentified places. This rectifying transformation is
\begin{equation} \label{eqn:trafo}
\begin{aligned}
\bar{\lambda}_i &= m_{\Lambda} \, \Lambda_i + \lambda_{0k} \\
\bar{\phi}_i &= m_{\Phi} \, \Phi_i + \phi_{0k} \; ,
\end{aligned}
\end{equation}
where $m_{\Lambda}$, $m_{\Phi}$, $\lambda_0$, and $\phi_0$ are parameters derived from an inversion of the transformation (\ref{eqn:beogl}):
\begin{align}
m_{\Lambda} &= 1/{m_{\lambda}} \; , \; \; m_{\Phi} =  1/{m_{\phi}} \\
\lambda_{0k} &= - \Lambda_{0k}/{m_{\lambda}} \; , \; \; \phi_{0k} = -  \Phi_{0k}/{m_{\phi}} \,.
\end{align}
The analysis method and the determination of new identifications are applied repeatedly.

\section{England} \label{sec:england}

Initially, the analysis of the distortions of Ptolemaic England (GH II.3) by Strang \citep{str97,str98b} is discussed.
From the south to the north of England, Strang identifies five regions having differing rotations (absolute value of the rotation angles $\leq$20\degree) and a common pivot point at Long Melford.
The procedure of his determination of these regions is the following.
The latitudinal scale of Ptolemy's positions is assumed to be 62.5 Roman miles (R.mi.) per 1\degree\ (actually $1\degree \mathrel{\widehat{=}} 111\,\mathrm{km} = 75\,\mathrm{R.mi.}$).
The longitudinal scale is determined on the basis of selected places by a comparison of Ptolemaic longitudinal distances with those in a modern map.
The result is 41.67\,R.mi./\degree.
Based on the assumed scales, the Ptolemaic positions are plotted and superimposed on a modern map with coincidence at \textit{Londinium}/London.
For selected places, the residual vector between the Ptolemaic position and the respective position on the modern map is drawn.
From the orthogonal bisectors of the residual vectors is expected that they meet in the pivot point.
Strang \citep[Fig. 6]{str97} presents a map with residual vectors and orthogonal bisectors for five places: \textit{Itunae aestuarium}/mouth of the Eden (No. 17 in Kleineberg et al. \citep{kle12}, which is the reference for the numbering in the following), \textit{Ganganorum promontorium}/Braich-y-Pwll (No. 23), \textit{Tamarus fluvius}/mouth of the Tamar (No. 35), \textit{Vedra fluvius}/mouth of the Wear (No. 56), \textit{Maridunum}/Carmarthen (No. 108).

The method described was reapplied in order to get an insight into its reliability. In doing so, formula (\ref{eqn:trafo}) was applied for a centring with respect to London ($\Lambda=20\degree$, $\Phi=54\degree$) and for a scaling.
$m_\Phi$ is $62.5\,\mathrm{R.mi.}/75\,\mathrm{R.mi.}\approx0.83333$. The longitudinal scale of 41.67\,R.mi.$/$\degree\ is assumed to be valid here for the south of England, i.e. for $\phi=50\degree$, so that $m_\Lambda=41.67\,\mathrm{R.mi.}/(75\,\mathrm{R.mi.}\cos{50\degree})\approx0.86436$.
The residual vectors were computed for 50 known places consentaneously identified by Rivet and Smith \citep{riv79}, Strang \citep{str98a}, and Kleineberg et al. \citep{kle12} (Nos. 17, 20, 21, 23, 25--27, 29, 31, 33, 35--37, 41, 56, 58, 59, 61, 64, 65, 88--91, 94--97, 99--102, 104--119, 123, 135).
For the Ptolemaic coordinates the values of the $\Omega$-recension and of the X-manuscript were used separately with the exception of No. 108, for which $\Phi=55\degree$ given by Nobbe \citep{nob43} was used.
As a result, the directions of the residual vectors and their orthogonal bisectors strongly depend on the parameters used.
Fig. \ref{fig:strang} shows the residual vectors based on $\Omega$.
The vectors of Nos. 17, 23, 35, 56, and 108 are in acceptable agreement with those shown by Strang \citep{str97}.
However, obviously the orthogonal bisectors do not meet in the alleged pivot point at Long Melford in general, not even if the parameters used are modified.
Accordingly, the results of Strang \citep{str97} are doubtful.
At last, the three places \textit{Eboracum Legio VI Victrix}/York (No. 94), \textit{Isurium}/Aldborough (No. 91), and \textit{Caturactonium}/Catterick (No. 89) are considered exemplarily. They are actually located towards the north-west but in the GH towards the north (cf. Fig. \ref{fig:strang}), which indicates a rotation.
However, the places have $\Lambda=20\degree$ so that they are rather roughly positioned than rotated.

A new investigation of the Ptolemaic places of \textit{Albion} was carried out interdisciplinarily, whereby the analysis method described in Section \ref{sec:methode} was applied; the results are given in Kleineberg et al. \citep[p. 35 ff.]{kle12}.
The distortions of the Ptolemaic places in England could be described satisfactorily by shifts of groups of places and longitudinal and latitudinal scalings.
The scale factors $m_\lambda=1.35$ and $m_\phi=1.30$ given by Kleineberg et al. \citep[p. 203 f.]{kle12} for \textit{Albion} are based on a deficient model for the distortions of Scotland; however, a recalculation only for England yields similar results: $m_\lambda = 1.379 \pm 0.037 $, $m_\phi = 1.363 \pm 0.044$.
Accordingly, an identical scale factor of ca. 1.35 can be assumed for longitude and latitude.
Based on these values, a reapplication of the analysis method resulted in no significant changes.

Nine transformation units Al8--16 were determined for England. Table \ref{tab:eng_trdiff} gives their relative shifts with respect to the central transformation unit Al14 ($\Lambda_{0\mathrm{R}} = 20\degree32'$, $\Phi_{0\mathrm{R}} = -15\degree13'$) based on formula (\ref{eqn:trdiff}).
The relative shifts are shown in Fig. \ref{fig:eng_trdiff} (\textit{Mona insula}/Man from \textit{Hibernia}, GH II.2, is assigned to Al10).
For this plot in the modern reference system the relative shifts were scale-corrected by $\Delta\Lambda_{0k}/m_{\lambda}$ and $\Delta\Phi_{0k}/m_{\phi}$.
The latitudinal relative shifts are similar and maximally ca. $\frac{1}{2}\degree$; the longitudinal relative shifts are nearly twice as large (in km).
Hence, the shape of Ptolemaic England is more accurate in latitude than in longitude.

\section{Scotland} \label{sec:schottland}

The places of Ptolemaic Scotland and nearby islands are given in Table \ref{tab:ortsliste} (\textit{ae.} = \textit{aestuarium}/estuary; \textit{fl.} = \textit{fluvius}/river, in Scotland all positions of rivers refer to river mouths; \textit{pr.} = \textit{promontorium}/cape or foreland).
Ptolemy divides his description of \textit{Albion} into the 'north side' (Scotland Nos. 1--11), the 'west side' (Scotland Nos. 12--17), the 'south side', the 'east and south side' (Scotland Nos. 42--55), 'towns' (Scotland Nos. 66--86), and 'islands' (Scotland Nos. 125--132).
Among the latter, five points (Nos. 128--132) describe the position and shape of \textit{Thule insula}.
The place \textit{Alauna} (No. 79) does not occur in the $\Omega$-recension but in the X-manuscript.

Fig. \ref{fig:albion} shows the Ptolemaic positions of \textit{Albion}. Obviously, the northern part is turned to the east.
It begins at Hadrian's Wall located between Solway Firth and Newcastle upon Tyne, and, accordingly, almost corresponds to modern Scotland.

\subsection{Objective of the turning of Scotland} \label{sec:ursprung}

Jones and Keillar \citep{jon96} assume an arithmetical origin for the turning of Ptolemaic Scotland because even Ireland (\textit{Hibernia} in GH II.2), which was not a part of the Roman Empire, was described accurately by Ptolemy and the island \textit{Ebuda}/Inner Hebrides (Islay according to \citep[p. 31]{kle12}) nearby Scotland, for example, is located relatively correctly in relation to the north of Ireland.
The high accuracy of the Ptolemaic positions of \textit{Hibernia} is confirmed by an analysis according to the methods described in Section \ref{sec:methode}, see Kleineberg et al. \citep[p. 24 ff.]{kle12}.

The reason for an intentional turning of Scotland by Ptolemy is surely the position of \textit{Thule}.
Ptolemy gives the latitude 63\degree\ for the centre of \textit{Thule}.
In GH I.7.1 Ptolemy says that Marinos located \textit{Thule} at 63\degree\ latitude (cf. \citep[p. 69]{stu06}) so that, obviously, Ptolemy adopted this latitude from Marinos (also suggested by e.g. \citep[pp. 73, 77]{dil85}).
In antiquity \textit{Thule} constituted the northern limit of the \textit{Oikoumene} (cf. \citep[p. 69, note 33]{stu06}) so that Ptolemy was obliged to arrange \textit{Albion} south of \textit{Thule}.
Because of too large latitudes in the southern part of Great Britain (e.g. London: $\phi=51\degree30'$, $\Phi=54\degree$) and the latitudinal scaling ($m_\phi>1$), the problem occurred that there was not enough space for the northern part of \textit{Albion} south of \textit{Thule}.
A way out was the turning of the northern part of \textit{Albion} into the free space in the east.
(Dilke \citep{dil84} assumes that a bending to the east is due to the traditional triangular shape of \textit{Albion} given by Eratosthenes; see also \citet{tie59} in this regard.)
In particular a rotation was suitable because, properly performed, it does not change the distances of the places of the rotated part.
Hadrian's Wall, the northern border of the Roman province \textit{Britannia}, lent itself to the limit of the turning.

Ptolemy's \textit{Thule} must be distinguished from Pytheas' \textit{Thule}, which presumably corresponds to the region of Trondheim in Norway (e.g. Hennig \citep[p. 168]{hen44}).
Ptolemy's \textit{Thule} is to be equated with the Shetland Islands (cf. Rivet and Smith \citep[p. 146]{riv79}, Dilke \citep[pp. 83, 136]{dil85}).
A reason for this is that according to Tacitus' \textit{Agricola} 10 the Romans named an archipelago \textit{Thule} which came within the range of vision during their circumnavigation of Great Britain, and this archipelago was surely the Shetland Islands.
{That information was certainly known to and used by Ptolemy.
Furthermore, in his \textit{Mathematike Syntaxis} (MS) II.6 Ptolemy assigns \textit{Thule} to the parallel at $\Phi=63\degree$ and Scythian people further north at $\Phi=64\degree30'$.
If Ptolemy referred to Pytheas' \textit{Thule}, the northern limit of the \textit{Oikoumene}, he would not locate people north of it.

The \textit{Orcades insulae} (No. 127), in the south of \textit{Thule}, are identified as the Orkney Islands (Rivet and Smith \citep[p. 433 f.]{riv79}).
The latitudinal distance of 1\degree\ between the southern point of \textit{Thule} (62\degree40$'$) and the \textit{Orcades insulae} (61\degree40$'$) is in good agreement with the actual distance of 50$'$ (from the southern tip of Shetland at $\lambda=-1\degree20$, $\phi=59\degree50'$ to the centre of Orkney at $\lambda=-3\degree00'$, $\phi=59\degree00'$), also if a scaling exists (e.g. factor 1.35: $1\degree/1.35=44'$).
Likewise, the longitudinal distance of 1\degree40$'$ between the easternmost point of \textit{Thule} (31\degree40$'$) and the \textit{Orcades insulae} (30\degree) coincides with the actual distance of 1\degree40$'$ (points as above).
Since the relative position of \textit{Thule} and the \textit{Orcades insulae} is correct, the \textit{Orcades insulae} were, obviously, not turned to the east together with Scotland so that its latitude was a further limit for the latitudinal dimension of \textit{Albion}.

As it can be seen in Fig. \ref{fig:albion}, the rotation of Scotland amounts to about 90\degree\ (with respect to the actual situation).
Richmond \citep{ric22} assumes a rotation around \textit{Vedra fl.}/Wear (No. 56) by 90\degree\ and performs a rotational correction of the Ptolemaic positions by means of an exchange of $\Lambda$ and $\Phi$.
Strang \citep{str97} also presumes a rotation around \textit{Vedra fl.} and determines a rotation angle of 20\degree\ for the north of England and of 70\degree\ for Scotland with respect to the north of England, together 90\degree.
Furthermore, he determines an additional rotation of places in the south of Scotland.
Rivet and Smith \citep[p. 114]{riv79} assume a rotation around \textit{Itunae aestuarium}/Eden (No. 17) by ca. 50\degree.
The reason for this rotation angle is the cape \textit{Epidium pr.}/Mull of Kintyre (No. 6), which appears in the description of \textit{Hibernia} as the island \textit{Epidium}.
The assumed rotation makes both places coincide. The equation of both places, however, is not mandatory.
By Kleineberg et al. \citep[p. 32]{kle12} the island \textit{Epidium} is identified as Arran.
Furthermore, the rotation-corrected position of \textit{Epidium pr.} does not need to coincide with the Ptolemaic position of \textit{Epidium} if Ptolemy adjusted the positions of \textit{Hibernia} to the afore rotated positions of Scotland.

\subsection{Adjustment model for the turning of Scotland} \label{sec:ausgleichung}

Supposing an intentional rotation for Ptolemaic Scotland, an accurate way for its accomplishment would have been a 3D rotation of points on the earth surface performed by means of a rotation around an axis through the point of origin and a given pivot point.
In the MS Ptolemy describes problems of the spherical astronomy, in MS VIII.5 the conversion of ecliptic longitude and latitude into right ascension and declination (see \citep[p. 97 ff.]{ped11}).
This conversion can be achieved by a rotation of the ecliptic or equatorial, respectively, reference system around the first coordinate axis through the vernal equinox by the angle of the obliquity of the ecliptic.
A similar problem is a 3D rotation of points around an arbitrary axis, which can be solved by at least three single rotations around coordinate axes.
Accordingly, it is imaginable that Ptolemy was able to perform a 3D rotation around an axis (pivot point) by means of a decomposition into single rotations around coordinate axes and that he applied the procedure to the places of Scotland or to some selected places.
That gave reason to model the turning of Scotland by a 3D rotation.
But even if Ptolemy proceeded in another way, e.g. by rotating mapped points computationally or graphically in the plane, a 3D rotation is a good approximation for the unknown original procedure.

The rotation applied to Scotland is an anti-clockwise rotation around an axis through the origin of the Ptolemaic coordinate system and the pivot point at $\Lambda_\mathrm{P}$, $\Phi_\mathrm{P}$ by an angle $\alpha$.
The direction of the axis is given by the unit vector
\begin{equation}
\mathbf{p} =
\begin{pmatrix}
p_1\\
p_2\\
p_3
\end{pmatrix}
=
\begin{pmatrix}
\cos\Lambda_\mathrm{P} \cos\Phi_\mathrm{P} \\
\sin\Lambda_\mathrm{P} \cos\Phi_\mathrm{P} \\
\sin\Phi_\mathrm{P}
\end{pmatrix} \;.
\end{equation}
The geometric transformation of the 3D position vector $\mathbf{x}_i$ of a place into its rotated position vector $\mathbf{X}_i$ by means of the mentioned rotation is
\begin{equation} \label{eqn:rotation}
\mathbf{X}_i = \mathbf{R}(\Lambda_\mathrm{P}, \Phi_\mathrm{P}, \alpha) \; \mathbf{x}_i \; ,
\end{equation}
where $\mathbf{R}$ is a rotation matrix.
$\mathbf{R}$ can be based on different compositions of single rotations around coordinate axes.
A description based on the elements of $\mathbf{p}$ is (cf. Bronstein et al. \citep[p. 301]{bro08}):
\begin{equation} \label{eqn:rotmatrix}
\mathbf{R} =
\begin{pmatrix}
p_1^2(1-\cos\alpha)+\cos\alpha &
p_1p_2(1-\cos\alpha)-p_3\sin\alpha &
p_1p_3(1-\cos\alpha)+p_2\sin\alpha \\
p_1p_2(1-\cos\alpha)+p_3\sin\alpha &
p_2^2(1-\cos\alpha)+\cos\alpha &
p_2p_3(1-\cos\alpha)-p_1\sin\alpha \\
p_1p_3(1-\cos\alpha)-p_2\sin\alpha &
p_2p_3(1-\cos\alpha)+p_1\sin\alpha &
p_3^2(1-\cos\alpha)+\cos\alpha
\end{pmatrix} \;.
\end{equation}

The objective is to estimate the unknowns $\Lambda_\mathrm{P}$, $\Phi_\mathrm{P}$, $\alpha$ by means of a least-squares adjustment on the basis of control points.
The observations of the adjustment model are the Ptolemaic coordinates $\Lambda_i$ and $\Phi_i$, which are composed to the observation vector
\begin{equation}
\mathbf{l} = (\ldots \Lambda_i \; \Phi_i \ldots)^\top \; .
\end{equation}
Using the unit sphere, the position vector $\mathbf{X}_i$ expressed by $\Lambda_i$ and $\Phi_i$ is
\begin{align}
\begin{pmatrix}
X_{i1}\\
X_{i2}\\
X_{i3}
\end{pmatrix}
&=
\begin{pmatrix}
\cos(\Lambda_i+v_{\Lambda\,i}) \cos(\Phi_i+v_{\Phi\,i}) \\
\sin(\Lambda_i+v_{\Lambda\,i}) \cos(\Phi_i+v_{\Phi\,i}) \\
\sin(\Phi_i+v_{\Phi\,i})
\end{pmatrix} \; ,
\end{align}
wherein the corrections $v_{\Lambda\,i}$, $v_{\Phi\,i}$ for the observations are introduced.
The position vector $\mathbf{x}_i$ is expressed by means of the presumable ancient longitude and latitude before the rotation.
To them the distortion model (\ref{eqn:beogl}) is applied so that longitude and latitude are replaced by $m_\lambda \lambda_i + \Lambda_0$ and $m_\phi \phi_i + \Phi_0$:
\begin{align} \label{eqn:x}
\begin{pmatrix}
x_{i1}\\
x_{i2}\\
x_{i3}
\end{pmatrix}
&=
\begin{pmatrix}
\cos(m_\lambda \lambda_i + \Lambda_0) \cos(m_\phi \phi_i + \Phi_0) \\
\sin(m_\lambda \lambda_i + \Lambda_0) \cos(m_\phi \phi_i + \Phi_0) \\
\sin(m_\phi \phi_i + \Phi_0)
\end{pmatrix}
\; .
\end{align}
The (unknown) differences in the shifts of groups of places must be neglected and average shift parameters $\Lambda_0$ and $\Phi_0$ must be used.
They are additional unknowns of the adjustment model so that the
vector of unknowns becomes:
\begin{equation}
\mathbf{u} = (\Lambda_\mathrm{P} \; \Phi_\mathrm{P} \; \alpha \; \Lambda_0 \; \Phi_0)^\top \; .
\end{equation}

The modern coordinates $\lambda_i$, $\phi_i$ are constants in the model as well as the scale factors $m_\lambda$, $m_\phi$, which are set at postulated values (resulting from the investigation of Ptolemaic England).

A rearrangement of formula (\ref{eqn:rotation}) yields the condition equations
\begin{align}
\mathbf{f}_i &=
 \mathbf{R}(\Lambda_\mathrm{P}, \Phi_\mathrm{P}, \alpha) \; \mathbf{x}_i (\Lambda_0, \Phi_0) - \mathbf{X}_i(\Lambda_i, \Phi_i) = \mathbf{0}
\end{align}
for $i=1(1)n$, where $n$ is the number of control points.
Because of the dependencies of the three components of $\mathbf{f}_i=(f_{i1}\; f_{i2}\; f_{i3})^\top$, two of them are sufficient for the system of condition equations of the adjustment.
The two equations
\begin{equation}  \label{eqn:bedgl}
\begin{aligned}
\psi_{i1} &= f_{i2}(\Lambda_i, \Phi_i, \Lambda_\mathrm{P}, \Phi_\mathrm{P}, \alpha, \Lambda_0, \Phi_0) = 0 \\
\psi_{i2} &= f_{i3}(\Phi_i, \Lambda_\mathrm{P}, \Phi_\mathrm{P}, \alpha, \Lambda_0, \Phi_0) = 0
\end{aligned}
\end{equation}
are chosen so that $\psi_{i1}$ and $\psi_{i2}$ are composed for each control point.

The equations (\ref{eqn:bedgl}) lead to an adjustment of nonlinear condition equations with unknowns (see e.g. \citep{nei10}).
The condition equations must be linearised and the unknowns must be determined iteratively.
The linearisation is based on approximate values $\bar{\mathbf{l}}^0 = \mathbf{l} + \mathbf{v}^0$ and $\mathbf{u}^0$:
\begin{align}
\mathbf{B} (\bar{\mathbf{l}} - \bar{\mathbf{l}}^0) + \mathbf{A} (\mathbf{u}-\mathbf{u}^0) + \psi(\bar{\mathbf{l}}^0, \mathbf{u}^0) &= \mathbf{0} \; ,
\end{align}
where $\bar{\mathbf{l}}$ are the adjusted observations and $\mathbf{\psi}=(\ldots \psi_{i1} \; \psi_{i2} \ldots)^\top$.
The matrices $\mathbf{B}$ and $\mathbf{A}$ contain the partial derivatives of $\mathbf{\psi}_{i}$ with respect to the observations and unknowns, respectively, which are computed on the basis of the approximate values.
By means of $\bar{\mathbf{l}} - \bar{\mathbf{l}}^0 = \bar{\mathbf{l}} - \mathbf{l} - \mathbf{v}^0 = \mathbf{v} - \mathbf{v}^0$, the common condition equations
\begin{align} \label{eqn:bedgl_lin}
&\mathbf{B} \mathbf{v} + \mathbf{A} (\mathbf{u}-\mathbf{u}^0) - \mathbf{w} = \mathbf{0} \\
&\mathbf{w} = \mathbf{B} \mathbf{v}^0  - \mathbf{\psi}(\bar{\mathbf{l}}^0, \mathbf{u}^0)
\end{align}
of a linear Gau\ss{}-Helmert-model are obtained, where $\mathbf{w}$ is the vector of misclosures.

The minimisation of the objective function $\mathbf{v}^\top\mathbf{P}\mathbf{v}$ with the side conditions (\ref{eqn:bedgl_lin}) is carried out as usual by means of the Lagrange multipliers (see e.g. \citep[p. 156 ff.]{nie02}).
$\mathbf{P}$ is the weight matrix of the stochastic part
\begin{equation}
\mathbf{C}_l = \sigma_0^2 \mathbf{P}^{-1}
\end{equation}
of the adjustment model, where $\mathbf{C}_l$ is the covariance matrix of the observations and $\sigma_0^2$ is the variance of unit weight. Possible correlations are not considered because they are unknown.
Since (unknown) relative shifts of places cannot be taken into account in the functional model (\ref{eqn:bedgl}), possible shifts are modelled by larger standard deviations in $\mathbf{C}_l$.

The unknowns are in part highly correlated (a change in $\Lambda_\mathrm{P}$, $\Phi_\mathrm{P}$ is compensated by $\Lambda_0$, $\Phi_0$, $\alpha$), which leads to large uncertainties of the adjusted unknowns.
A way out is an `adjustment with stochastic advance information' (see e.g. \citep[p. 240 f.]{nie02}, \citep[p. 167 ff.]{bau93}), in which additional pseudo-observations are introduced for the unknowns.
According to this, the observations $l_{\Lambda_0}$ and $l_{\Phi_0}$ are added for $\Lambda_0$ and $\Phi_0$ because for them advance information is available from the investigation of England.
In addition to the condition equations (\ref{eqn:bedgl}), the equations
\begin{equation} \label{eqn:bedgl2}
\begin{aligned}
l_{\Lambda_0} + v_{\Lambda_0} - \Lambda_0 &= 0\\
l_{\Phi_0} + v_{\Phi_0} - \Phi_0 &= 0
\end{aligned}
\end{equation}
appear.

In a further adjustment $\alpha$ is not an unknown but is set at a constant value.

\subsection{The rotation of Scotland} \label{sec:drehung}

The adjustment described in the last section was applied to:
1) the identifications for the Ptolemaic places given by Rivet and Smith \citep[p. 237 ff.]{riv79} in conjunction with the ancient coordinates a) of $\Omega$ and b) of X; 2) the identifications given by Strang \citep{str98a} in conjunction with the ancient coordinates a) of $\Omega$ and b) of X; 3) the identifications and ancient coordinate variants determined by the present work (cf. Section \ref{sec:analyse}).

Firstly, the computations 1) and 2) are considered. In the case of 1) no identifications are given for the places Nos. 8, 73, 83.
Table \ref{tab:rotpar} gives the a priori standard deviations $\sigma_{\Lambda i}$ and $\sigma_{\Phi i}$ of the observations $\Lambda_i$ and $\Phi_i$ and the standard deviation of unit weight $s_0$ from the adjustment; its a priori value is $\sigma_0=1$.
$\sigma_{\Lambda i}$ and $\sigma_{\Phi i}$ were chosen such that the overall model test (significance level 5\%) showed no errors, i.e. $s_0 \approx \sigma_0$.
For the constants $m_\lambda$ and $m_\phi$ the value 1.35 was adopted from England (see Section \ref{sec:england}).
For the observations $l_{\Lambda_0}$ and $l_{\Phi_0}$ of the unknowns $\Lambda_0$ and $\Phi_0$, the results $\Lambda_{0\,8}=21\degree15'$, $\Phi_{0\,8}=-15\degree28'$ of the adjustment of the northernmost transformation unit Al8 in England (cf. Fig. \ref{fig:eng_trdiff}) are first guidelines.
However, preceding investigations of Scotland turned out that the majority of the Scottish places are shifted by ca. 1\degree\ further towards the north on average (see Section \ref{sec:analyse}) so that $l_{\Phi_0}=-15\degree28'+1\degree=-14\degree28'$ was used.
The standard deviations used are $\sigma_{\Lambda_0}=30'$ and $\sigma_{\Phi_0}=15'$.

The results of the adjustment are given in Table \ref{tab:rotpar}.
Fig. \ref{fig:ell} shows the adjusted pivot points P$_1$ and P$_2$ for the computations 1.b) and 2.b), respectively, and the places in the vicinity (X-coordinates).
The geographic coordinates were regarded as two-dimensional coordinates, and the confidence ellipses of P$_1$ and P$_2$ were computed from the cofactor matrix of the unknowns and $s_0$ as usual in the adjustment of geodetic networks (e.g. \citep[p. 258]{nie02}).
Fig. \ref{fig:ell} shows the ellipses based on a probability of 80\%.

Taking into account the uncertainty of the results, the estimated rotation parameters are $\Lambda_\mathrm{P}\approx18\degree$, $\Phi_\mathrm{P}\approx58\degree30'$, and $\alpha\approx-80\degree$ in each case.
The rotation parameters in conjunction with $\Lambda_0$ and $\Phi_0$ apply to a transformation between the Ptolemaic and modern coordinates.
Taken alone, the rotation parameters are also approximations for the transformation between the Ptolemaic and the not rotated ancient coordinates if Scotland was oriented correctly before its rotation.
In the following, possible original rotation parameters are discussed.

For the pivot point, firstly places of the GH are taken into account which are located near the adjustment result.
Assuming that \textit{Itunae ae.}/River Eden (No. 17; cf. \citep{tho75}, \citep[p. 380]{riv79}) is rotated, its rotation-corrected position is used as a criterion.
For the rotation angle, 80\degree--90\degree\ are presumed according to the result of the adjustment.
The results for the considered places (cf. Fig. \ref{fig:ell}) are:
pivot point \textit{Trimontium} (No. 71): \textit{Itunae ae.} (No. 17) is too far east;
pivot point \textit{Vinovium} (No. 88): \textit{Itunae ae.} (No. 17) is too far west;
pivot point \textit{Epiacum} (No. 87): \textit{Itunae ae.} (No. 17) is too far south with respect to \textit{Moricambe ae.}/Morecambe Bay (No. 18; \citep[p. 43]{kle12}) and \textit{Epiacum}/Wreay (No. 87; \citep[p. 51]{kle12});
pivot point \textit{Itunae ae.} (No. 17): \textit{Itunae ae.} is too far east with respect to \textit{Moricambe ae.} (No. 18) and \textit{Epiacum} (No. 87);
pivot point \textit{Moricambe ae.} (No. 18): \textit{Itunae ae.} (No. 17) is positioned without disagreement.
A somewhat better result than by means of \textit{Moricambe ae.} (No. 18) is achieved by the point P$_\mathrm{P}=(\Lambda=18\degree$, $\Phi=58\degree30'$), which corresponds to the result of the adjustment.
By means of this point, \textit{Itunae ae.} (No. 17) is positioned correctly north of \textit{Moricambe ae.} (No. 18).
Moreover, the round coordinate values of P$_\mathrm{P}$ argue for this pivot point because they are easy to handle and possibly eased the procedure of the rotation (e.g. a calculation).
Nonetheless, no certain conclusion on the exact position of the pivot point can be drawn here owing to the inaccuracy of the ancient coordinates; due to the arguments for P$_\mathrm{P}$, however, this point is assumed to be the pivot point in the following.

A rotation angle of 90\degree\ is less probable because the correction of this rotation leads to a northern direction of the east coast between \textit{Boderia aestuarium}/Firth of Forth (No. 54; \citep[p. 47]{kle12}) and \textit{Taezalorum pr.}/Kinnairds Head (No. 50; \citep[p. 46]{kle12}) in disagreement with the actual northeastern direction.
For this coast a more accurate direction can be expected, because the southern part of the east coast between \textit{Vedra fl.}/Wear (No. 57; \citep[p. 46]{kle12}) and \textit{Boderia aestuarium} has a correct northwestern direction.
Hence, a rotation angle $<90\degree$ is probable.
As in the case of the pivot point, a round figure comes into consideration, that is 80\degree, which is in agreement with the adjustment result.
Probably, Ptolemy chose a rotation by which \textit{Dumna insula} (No. 126) can be positioned south of the latitude of the not rotated \textit{Orcades insulae} (No. 127), cf. Fig. \ref{fig:albion} and Section \ref{sec:ursprung}.
Moreover, an angle of 80\degree\ in conjunction with P$_\mathrm{P}$ is probable because the correction of this rotation positions the four places \textit{Verubium} (No. 43), \textit{Ila fl.} (No. 44), \textit{Ripa alta} (No. 45), and \textit{Varar ae.} (No. 47; not $\Omega$-coordinates, only X) of the (actual) east coast almost exactly at 16\degree30$'$ longitude, cf. Table \ref{tab:ortsliste}, column $\Lambda^*$ and Fig. \ref{fig:coast_a}.
It is likely that the northeastern direction of this northern-most section of the east coast was not known to Ptolemy and that he positioned the places northwards at the same meridian.
For the mentioned reasons, the angle $\alpha_\mathrm{P}=-80\degree$ is assumed to be the rotation angle in the following.

For computation 3) the average shift parameters of Scotland were newly determined, see Section \ref{sec:analyse}.
The resulting parameters and standard deviations were used for the observations $l_{\Lambda_0}$ and $l_{\Phi_0}$.
$\alpha$ was set at $\alpha=\alpha_\mathrm{P}$ constantly.
The estimated pivot point P$_3$ of computation 3) is located nearby P$_\mathrm{P}$ (cf. Table \ref{tab:rotpar}).
Its confidence ellipse is significantly smaller than those of computations 1) and 2) and contains no place of the GH but only P$_\mathrm{P}$ (cf. Fig. \ref{fig:ell}).
Accordingly, this subsequent adjustment confirms the pivot point P$_\mathrm{P}$ and the angle $\alpha_\mathrm{P}$.

Fig. \ref{fig:bri_rot} shows the places of \textit{Albion} ($\Omega$-recension), wherein the places of Scotland are rotation-corrected by means of
\begin{equation} \label{eqn:korrektur_rot}
\begin{pmatrix}
\Lambda_i^* \\
\Phi_i^*
\end{pmatrix}
=
\mathbf{R}(\mathrm{P}_\mathrm{P}, \alpha_\mathrm{P})^{-1}
\begin{pmatrix}
\Lambda_i \\
\Phi_i
\end{pmatrix}
\; ,
\end{equation}
with the inverse matrix of $\mathbf{R}$ (formula (\ref{eqn:rotmatrix})).
The Scottish part is in good agreement with the actual shape of Scotland.
The latitudinal distance from \textit{Cantium pr.}/South-Foreland (No. 41; \citep[p. 45]{kle12}) at the south coast of Great Britain to \textit{Tarvedum sive Orcas pr.} (No. 11) at the north coast is $65\degree-54\degree=11\degree$, and the true latitudinal distance is ca. $7\degree30'$.
Taking into account the average relative latitudinal shift of Scotland of ca. 1\degree\ (see Section \ref{sec:analyse}) and the scale factor $m_\phi=1.35$, the ancient distance becomes $(11\degree-1\degree)/1.35=7\degree24'$, which only differs by 6$'$ from the actual distance.

The correction of Ptolemy's rotation yields an overlap of a few Scottish places with places from Ptolemy's description of \textit{Hibernia}.
That does not contradict the assumed rotation if Ptolemy compiled his description of \textit{Hibernia} after the rotation of Scotland.

In Book VIII of the GH positions of the so-called \textit{poleis episemoi} (noteworthy cities) are given.
Among the places in Ptolemaic Scotland and the northern islands, \textit{Pinnata castra} (No. 83),  \textit{Dumna insula} (No. 126), and \textit{Thule} (centre, No. 132) are listed in Book VIII.
The positions are expressed there by means of the time difference $A$ (in hours) from the location to \textit{Alexandria} and the length of the longest day $M$ (in hours) at the location.
The coordinates in Book VIII were presumably determined from the coordinates in the location catalogue; the $M$-data probably originate from a linear interpolation of a compilation of parallels with specific $M$ in MS II.6, see Marx \citep{mar12b}.
The $A$ of Nos. 83 and 126 are probably based on the longitude of \textit{Alexandria} $\Lambda_\mathrm{A}=60\degree$ (given in GH VIII.15.10), whereas $A$ of No. 132 is better explicable by $\Lambda_\mathrm{A}=60\degree30'$ (GH IV.5.9).
$\Phi$ (only $\Omega$) and $M$ of No. 126 correspond to those of the 27th parallel in MS II.6, and $\Phi$ and $M$ of No. 132 correspond to those of the 29th parallel.
No. 83 is situated between the 25th and 26th parallel.
Its $M=18^\mathrm{h}30^\mathrm{m}$ is possibly the result of a linear interpolation which yields $\approx18^\mathrm{h}27^\mathrm{m}$ so that $M$ of No. 83 is a rounded value.
$\Lambda$ and $\Phi$ of the three considered places are in good agreement with the coordinates of Book VIII.
That does not hold true, however, for the rotation-corrected coordinates $\Lambda^*$ and $\Phi^*$ (Table \ref{tab:ortsliste}) so that, obviously, the coordinates of Book VIII originate from Ptolemy's rotated coordinates of the location catalogue.

The 23rd parallel in MS II.6 is at 56\degree\ and `goes through the middle of Great Brittania' (Toomer \citep[p. 88]{too84}), Ptolemy's name for \textit{Albion} in the MS. Latitude 56\degree\ is in agreement with the latitudinal dimension of \textit{Albion} in the GH (cf. Fig. \ref{fig:albion}) but not with its rotation-corrected dimension (Fig. \ref{fig:bri_rot}).
Accordingly, even though the MS was written before the GH (Dilke \citep[p. 212, note 30]{dil85}), the rotation of Scotland was presumably performed before the preparation of MS II.6.

\subsection{Islands, waters and points at the coast} \label{sec:kueste}

The islands near Scotland given by Ptolemy are located only in the north and northeast of Ptolemaic Scotland  (cf. Fig. \ref{fig:albion}).
\textit{Dumna insula} (No. 126) and \textit{Orcades insulae} (No. 127) are located at $\Lambda=30\degree$, the centre of \textit{Thule} at $\Lambda=30\degree20'$.
On Ptolemy's localisation of the \textit{Orcades insulae} and \textit{Thule} see Section \ref{sec:ursprung}.
\textit{Dumna insula} was most likely rotated together with Scotland and can be identified as Lewis (see Rivet and Smith \citep[p. 342]{riv79} and Section \ref{sec:analyse}) in the east of Scotland.

The shape of \textit{Albion} gives rise to the question of whether the longitudinal dimension of Ptolemaic Scotland was adjusted to a predetermined longitude of \textit{Thule} or whether the longitude of \textit{Thule} was determined by the longitudinal dimension of Ptolemaic Scotland.
The latter is certainly the case because the longitudinal dimension of Ptolemaic Scotland corresponds to its latitudinal dimension before the rotation, which is, apart from further systematic errors, correct (cf. Section \ref{sec:drehung}).
Accordingly, accurate data sources can be assumed, by which the longitudinal dimension of the rotated Ptolemaic Scotland was determined.
Presumably, Ptolemy simply equated the longitude of the \textit{Orcades insulae} with that of \textit{Dumna insula} so that also the longitude of \textit{Thule} was given by the relative position of the \textit{Orcades insulae} and \textit{Thule} (cf. Section \ref{sec:ursprung}).

\textit{Scitis insula} is usually identified as Skye (e.g. Watson \citep{wat26}, Rivet and Smith \citep[p. 452]{riv79}).
Its odd Ptolemaic position is explicable by an error in the original latitude before the rotation, see Section \ref{sec:analyse}.

Fig. \ref{fig:coast_a} shows the rotation-corrected Scottish points and waters at the coast (formula (\ref{eqn:korrektur_rot}), coordinates see Table \ref{tab:ortsliste}), which are connected by straight lines indicating the shape of Ptolemaic Scotland.
In contrast, the actual coast is shown in Fig. \ref{fig:coast_m} together with the known and assumed modern counterparts (cf. Table \ref{tab:ortsliste}).
Some important identifications are considered in the following.

From \textit{Tarvedum sive Orcas pr.} (No. 11) can be expected that it was that part of Great Britain which was located nearest to the \textit{Orcades insulae} (cf. Rivet and Smith \citep[p. 115]{riv79}).
Therefore, it is usually identified as Dunnet Head at the north coast (e.g. Rivet and Smith \citep[p. 422]{riv79}, Strang \citep{str98a}).
This identification, however, is contradictory to the rotation of Scotland.
In Ptolemy's description, \textit{Tarvedum sive Orcas pr.} is the last, easternmost place of the north side so that it should correspond to the northwestern corner of Scotland, i.e. Cape Wrath (also suggested by Bradley \citep[p. 7]{bra84}).
It is unlikely that this characteristic place was unknown to Ptolemy because it was surely of importance for the ancient navigation.
Possibly, the name \textit{Orcas pr.} was defined by Ptolemy himself because the point is that corner of \textit{Albion} which is nearest to the position of \textit{Orcades insulae} in his description of \textit{Albion}.
Further evidence for the identification as Cape Wrath is provided by the consistency of the Ptolemaic coordinates of \textit{Tarvedum sive Orcas pr.} with those of other places in the northwest of Scotland, see Section \ref{sec:analyse}.

Since \textit{Tarvedum sive Orcas pr.} is equated with Cape Wrath, the three next places of the actual north and east coast \textit{Virvedrum pr.} (No. 42), \textit{Verubium pr.} (No. 43), and \textit{Ila fl.} (No. 44) can be identified as Dunnet Head, Duncansby Head (also suggested by Bradley \citep[p. 8]{bra84}), and River Wick, as it is indicated by the actual shape of the coast.

From the rotation of Scotland it follows that \textit{Nabarus fl.} (No. 10) is located at the west coast of Scotland.
However, \textit{Nabarus fl.} is usually identified as the River Naver at the Scottish north coast because of a presumable relation between the ancient and modern name (cf. Rivet and Smith \citep[p. 422]{riv79}).
An explanation for this seeming disagreement is given in the following section.

\textit{Novantarum chersonesus et pr.} (No. 1) is usually identified as Mull of Galloway (e.g. Rivet and Smith \citep[p. 426 f.]{riv79}).
Since, however, it is the first place in Ptolemy's description of the north side of \textit{Albion}, which is confirmed by its coordinates, it is possibly rather the northwestern tip (Corsewall Point) of Rhinns of Galloway than the southeastern tip (Mull of Galloway).

\subsection{Application of the analysis method} \label{sec:analyse}

The analysis method described in Section \ref{sec:methode} was applied to the rotation-corrected Ptolemaic coordinates $\Lambda^*_i$ and $\Phi^*_i$ (formula (\ref{eqn:korrektur_rot})).
In distortion model (\ref{eqn:beogl}) $\Lambda$ and $\Phi$ had to be replaced by $\Lambda^*$ and $\Phi^*$.
Before the rotation all considered ancient coordinate variants of a place were combined so that all possible ancient point variants were generated (e.g. in the case of two variants for $\Lambda$ and $\Phi$, four point variants are possible).
The $\Omega$-coordinates and differing variants from X, M\"uller \citep{mue83}, and Nobbe \citep{nob43} were used.
In addition to the identifications mentioned in Section \ref{sec:kueste}, those of Hazlitt \citep{haz51}, M\"uller \citep{mue83}, Rivet and Smith \citep{riv79}, Thomas \citep{tho75}, and Watson \citep{wat26} were taken into consideration (for a compilation see Kleineberg et al. \citep[p. 42 ff.]{kle12}).

The a priori standard deviations $\sigma_{\Lambda^*i}$ and $\sigma_{\Phi^*i}$ of $\Lambda^*_i$ and $\Phi^*_i$ were chosen on the basis of those resulting from the adjustment of the English places.
The smallest among them are 12$'$ for $\Lambda$ and 9$'$ for $\Phi$. Converting 12$'$ from the mean latitude of England to the mean latitude of Scotland yields $13.4'$ so that $\sigma_{\Lambda^*i}=13'$, $\sigma_{\Phi^*i}=9'$ were applied.
In the cases of seemingly rough coordinate values, larger values were used: $\sigma_{\Lambda^*i}=18'$ for $\Lambda^*$ of No. 13 ($\Phi=61\degree$), No. 15 ($\Phi=60\degree$), No. 71 ($\Phi=59\degree$); $\sigma_{\Phi^*i}=12'$ for $\Phi^*$ of No. 14 ($\Lambda=19\degree$), No. 71 ($\Lambda=19\degree$).

The identifications determined by the analysis method and their transformation units are given in Table \ref{tab:ortsliste}.
For a few places no identification and/or transformation unit are given because either the modern or the ancient coordinates turned out to be inconsistent.
Additionally, Fig. \ref{fig:karte} shows a modern map of the Ptolemaic places including the transformation units.
Column 'SI' of Table \ref{tab:ortsliste} contains the sources of the identifications, they are: 'B': Bradley \citep{bra84}, 'H': Hazlitt \citep{haz51}, 'M': M\"uller \citep{mue83}, 'R': Rivet and Smith \citep{riv79}, 'T': Thomas \citep{tho75}.
Column 'SC' gives the sources of the determined ancient coordinate variants; in addition to $\Omega$ and X there is only 'M' for M\"uller \citep{mue83} in the case of No. 73 ($\Lambda=21\degree20'$).
In columns '$\Delta \lambda$' and '$\Delta \phi$' the residuals
\begin{equation}  \label{eqn:restf}
\begin{aligned}
\Delta \lambda_i &= \bar{\lambda}_i - \lambda_i \\
\Delta \phi_i &= \bar{\phi}_i - \phi_i
\end{aligned}
\end{equation}
after the rectifying transformation (\ref{eqn:trafo}) are given (based on $\Lambda^*_i$, $\Phi^*_i$).

Table \ref{tab:sch_trdiff} gives the relative shifts with respect to transformation unit Al7 (formula (\ref{eqn:trdiff})) with $\Lambda_{0\,7}=21\degree12'$ and $\Phi_{0\,7}=-15\degree27'$.
Fig. \ref{fig:sch_trdiff} shows the scale-corrected relative shifts $\Delta\Lambda_{0k}/m_{\lambda}$ and $\Delta\Phi_{0k}/m_{\phi}$ in the modern reference system.
The relative shifts are lower than 2\degree.
The latitudinal shifts with respect to Al7 are systematically northwards, on average ca. 1\degree.
Al7 is shifted with respect to the neighbouring Al8 in Ptolemaic England significantly only in longitude ($\Lambda_{0\,8}=22\degree15'$ and $\Phi_{0\,8}=-15\degree28'$).

In Table \ref{tab:sch_trdiff} the a posteriori standard deviations of the ancient coordinates resulting from the adjustment of the transformation units are given. They are scale-corrected by means of
\begin{equation} \label{eqn:stdabw}
\begin{aligned}
s_{\lambda^*i} &= s_{\Lambda^*i}/m_\lambda \\
s_{\phi^*i} &= s_{\Phi^*i}/m_\phi \; ;
\end{aligned}
\end{equation}
the few coordinates with larger a priori standard deviations (see above) were not involved.
Neglecting possibly underestimated values, the uncertainty is about 9--20\,km and corresponds to that in England.

On the basis of the 45 places assigned to transformation units, the scale factors were adjusted (using model (\ref{eqn:beogl})).
The results are $m_\lambda=1.38 \pm 0.06$ and $m_\phi=1.32 \pm 0.08$ based on the a priori standard deviations and $m_\lambda=1.34 \pm 0.04$ and $m_\phi=1.38 \pm 0.07$ based on the a posteriori standard deviations, which do not differ significantly from the postulated value 1.35.

Following the rating given by Kleineberg et al. \citep[p. 37 ff.]{kle12}, at least twelve places can be considered to be surely identified:  Nos. 15, 16, 17, 47, 48, 49, 51, 54, 71, 72, 80, 127.
Four of them, Nos. 47, 48, 49, 51, are in transformation unit Al3 in the northern part of Scotland.
The estimated uncertainties in Al3 are $s_{\lambda^*i} = 9'$ and $s_{\phi^*i} = 5'$ (formula (\ref{eqn:stdabw})) so that even this northern part turned out to be accurately determined.

\textit{Nabarus fl.} (No. 10) is consistent in transformation unit Al2 with its identification River Naver at the north coast.
According to the rotation of Scotland, however, it should be at the west coast.
This discrepancy is explicable by the different shifts of the northwestern places in Al1 and the northeastern places in Al2 (rotation-corrected situation, cf. Fig. \ref{fig:sch_trdiff}).
Possibly, Ptolemy had different data sources for these two regions, or his source already contained the significant relative shift of both regions.
Viewed from Al1, the places of Al2 are shifted in a west-southwestward direction.
That locates \textit{Nabarus fl.} at a longitude coinciding with the west coast (cf. Fig. \ref{fig:coast_a}).
Accordingly, Ptolemy assumed that \textit{Nabarus fl.} is located at the west coast, which yielded the position at the north coast in his rotated Scotland.

The assignment of \textit{Ripa Alta}/Tarbat Ness (No. 45) to Al1 is somewhat questionable because of its distant location at the east coast.
The assignment has been kept because the residuals in Al1 are very small and the identification is inconsistent in the nearest transformation unit Al3.
Possibly, the accurate position of \textit{Ripa Alta} with respect to the west coast originates from a common data source arisen from a circumnavigation.
Alternatively, \textit{Ripa Alta} could be identified as Ord of Caithness (according to Thomas \citep{tho75}), which is consistent in Al3.

The odd Ptolemaic position of \textit{Skitis insula} (No. 125) is explicable by means of an identification as Skye in conjunction with an error in the presumed original latitude of 66\degree\ ($\approx\Phi^*$) before the rotation.
Thereby, it is consistent in transformation unit Al1 with its rotation-corrected longitude $\Lambda^*=13\degree42'$ (possibly originally $13\degree40'$).
If, for example, the point $\lambda=-6\degree18'$, $\phi=57\degree42'$ at the north coast of Skye is chosen and the latitude is changed to 64\degree, an assignment to Al1 and adjustment (model (\ref{eqn:beogl}) with $m_\lambda=m_\phi=1.35$) indicate no error and the transformation (\ref{eqn:trafo}) yields the residuals $\Delta\lambda=-18'$, $\Delta\phi=0'$.
If Point of Sleat at $\lambda=-6\degree01'$, $\phi=57\degree01'$ at the south coast of Skye is chosen and the latitude is changed to $63\degree$, no errors is indicated and the residuals are $\Delta\lambda=-4'$, $\Delta\phi=3'$.

For the estimation of the pivot point (Section \ref{sec:drehung}) on the basis of the results of the present analysis, the average shift parameters of Scotland were of interest.
They were determined by means of model (\ref{eqn:beogl}), wherein the shifted groups of places were not taken into account.
The result based on the 45  places with assignment to a transformation unit is: $\Lambda_0= 21\degree30' \pm 5'$, $\Phi_0= -14\degree28' \pm 4'$.

\section{Summary and conclusion}

The turning of Ptolemy's places north of Hadrian's Wall (Ptolemaic Scotland) to the east was modelled by a 3D rotation, which turned out to describe the turning satisfactorily well.
The pivot point and the rotation angle $\alpha$ of a transformation between the Ptolemaic coordinates ($\Lambda_i$, $\Phi_i$) and modern coordinates ($\lambda_i$, $\phi_i$) were determined by means of adjustment theory on the basis of different data sets for the identifications of the Ptolemaic places.
From the results conclusions about the original rotation were derived.
The presumable pivot point was at (near) $\Lambda_\mathrm{P}=18\degree$, $\Phi_\mathrm{P}=58\degree30'$; the presumable rotation angle was (ca.) 80\degree.

Based on the resulting parameters, rotation-corrected Ptolemaic coordinates $\Lambda^*_i$ and $\Phi^*_i$ were computed.
The remaining differences between the $\Lambda^*_i$, $\Phi^*_i$ and $\lambda_i$, $\phi_i$ were modelled by scaling errors and shifts.
Groups of places of homogenous shifts (transformation units) in conjunction with best fitting modern counterparts of the Ptolemaic places were determined by means of a geodetic-statistical analysis method.
Based on the results, scale factors $m_\lambda$, $m_\phi$ of a transformation between the $\lambda_i$, $\phi_i$ and $\Lambda^*_i$, $\Phi^*_i$ were determined by means of an adjustment.
$m_\lambda$ and $m_\phi$ are ca. 1.35 in agreement with the result for the places south of Hadrian's Wall.
The factor $>1$ is caused by Ptolemy's underestimation of the circumference of the earth.
The transformation units have relative shifts $<$2\degree\ and coordinate accuracies of about 10--20\,km, which shows that Ptolemy had an extensive knowledge of Scotland, possibly owing to Roman military sources.

For the places of Great Britain there was not enough space in Ptolemy's description of the \textit{Oikoumene} owing to the scaling error and the preset latitude of \textit{Thule}.
Probably Ptolemy determined the geographic coordinates of the Scottish places or of some selected places first and then rotated them in order to satisfy the latitudinal limit given by \textit{Thule}.
From the position of the \textit{Orcades insulae}/Orkney north of Ptolemaic Scotland it can be deduced that it was not rotated.

Ptolemy's \textit{Thule} must be distinguished from Pytheas' \textit{Thule} and is to be equated with Shetland.
Reasons for this are the report of the sighting of \textit{Thule} by the Romans in Tacitus' \textit{Agricola}, the localisation of people north of \textit{Thule} in MS II.6, and the agreement of the relative position of \textit{Thule} and the \textit{Orcades insulae} with that of Shetland and Orkney.

Ptolemy's procedure in the determination of the positions of Scotland was presumably:
determining the latitude of the \textit{Orcades insulae} in the south of \textit{Thule};
choosing a pivot point and a rotation angle such that the rotated \textit{Dumna insula} is located south of the \textit{Orcades insulae};
determining the positions of Ptolemaic Scotland based on a rotation;
determining the longitudes of the \textit{Orcades insulae} and \textit{Thule} such that they are located north of the eastern end of the rotated Scotland.

Further main results concerning questions about Ptolemaic positions are the following.
The generally accepted equation of \textit{Nabarus fl.} with the River Naver seems to contradict the rotation of Scotland, since the Ptolemaic and the actual position are at the north coast.
That is explicable by two shifted regions in the north of Scotland, which led Ptolemy to assign it to the wrong coast side.
The characteristic, northwesternmost point of Scotland, Cape Wrath, is usually not equated with one of the Ptolemaic places.
It is, however, not missing in Ptolemy's description; it can be equated with \textit{Tarvedum sive Orcas pr.}, from which a position near to the \textit{Orcades insulae} can be expected.
That is not fulfilled by the real situation, but by Ptolemy's description of Scotland.

The identification of the Ptolemaic places was based on the ancient coordinates, the topographic situation, and the preliminary work of other authors.
An evaluation by other scientific disciplines is desirable.

\section*{Acknowledgement}

I thank Andreas Kleineberg for his collaboration on the compilation of the identifications of the Scottish Ptolemaic places to be found in the literature and for his information about the derivation of the place name \textit{Tarvedum sive Orcas promontorium}.

\clearpage

\begin{table} [hp]
\caption{Relative shifts of the transformation units (TU) in England with respect to Al14}\label{tab:eng_trdiff}
\centering
\begin{tabular}
{ l r r r r r r r }
\hline
\multicolumn{1}{c}{TU} & \multicolumn{1}{c}{$n$} & \multicolumn{1}{c}{$\Delta \Lambda_{0k}$} & \multicolumn{1}{c}{$\Delta \Phi_{0k}$} \\
\hline
Al8  &  6&    1\degree43$'$ & $-$0\degree16$'$ \\
Al9  &  6&    0\degree41$'$ & $-$0\degree36$'$ \\
Al10 & 18&    0\degree45$'$ & $-$0\degree05$'$ \\
Al11 &  4& $-$0\degree50$'$ & $-$0\degree33$'$ \\
Al12 &  3& $-$1\degree42$'$ & $-$0\degree25$'$ \\
Al13 &  9& $-$0\degree48$'$ &    0\degree10$'$ \\
Al14 & 10&    ---           &    ---           \\
Al15 & 13&    0\degree09$'$ & $-$0\degree32$'$ \\
Al16 &  4&    1\degree17$'$ & $-$0\degree30$'$ \\
\hline
\end{tabular}
\end{table}

{\setlength{\tabcolsep}{1.6mm}
\begin{longtable}[c]
{ >{\footnotesize{}}r >{\footnotesize{}\itshape}f{2.30cm} >{\footnotesize{}}l >{\footnotesize{}$}r<{$} >{\footnotesize{}$}r<{$} >{\footnotesize{}}f{2.30cm} >{\footnotesize{}}l *{2}{ >{\footnotesize{}$}r<{$} >{\footnotesize{}$}r<{$} } >{\footnotesize{}}c}
\caption{Identifications of the places in Ptolemaic Scotland}
\label{tab:ortsliste}\\
 \hline
 \multicolumn{1}{>{\footnotesize{}}c}{No.}
 & \multicolumn{1}{>{\footnotesize{}}c}{Ancient name}
 & \multicolumn{1}{>{\footnotesize{}}c}{SC}
 & \multicolumn{1}{>{\footnotesize{}}c}{$\Lambda^*$}
 & \multicolumn{1}{>{\footnotesize{}}c}{$\Phi^*$}
 & \multicolumn{1}{>{\footnotesize{}}c}{Identification}
 & \multicolumn{1}{>{\footnotesize{}}c}{SI}
 & \multicolumn{1}{>{\footnotesize{}}c}{$\lambda$}
 & \multicolumn{1}{>{\footnotesize{}}c}{$\phi$}
 & \multicolumn{1}{>{\footnotesize{}}c}{$\Delta\lambda$}
 & \multicolumn{1}{>{\footnotesize{}}c}{$\Delta\phi$}
 & \multicolumn{1}{>{\footnotesize{}}c}{TU} \\
 & & \multicolumn{1}{>{\footnotesize{}}c}{$\Lambda,\Phi$}
 & \multicolumn{1}{>{\footnotesize{}}c}{$[\degree,']$}
 & \multicolumn{1}{>{\footnotesize{}}c}{$[\degree,']$}
 & &
 & \multicolumn{1}{>{\footnotesize{}}c}{$[\degree,']$}
 & \multicolumn{1}{>{\footnotesize{}}c}{$[\degree,']$}
 & \multicolumn{1}{>{\footnotesize{}}c}{$[']$}
 & \multicolumn{1}{>{\footnotesize{}}c}{$[']$}
 &  \\
\hline
\endhead
1 & Novantarum chersonesus et pr. & $\Omega$,$\Omega$  &  12,08 &  60,20 & Corsewall Point (?) & --- & -5,10 &  55,00 & \text{---} & \text{---} & --- \\
2 & Rerigonius sinus & $\Omega$,$\Omega$  &  13,46 &  60,03 & Loch Ryan & R & -5,05 &  55,02 & 1 & 12 & Al6 \\
3 & Vindogara sinus & $\Omega$,$\Omega$  &  14,30 &  60,26 & Irvine Bay & R & -4,40 &  55,34 & 8 & -3 & Al6 \\
4 & Clota ae. & $\Omega$,$\Omega$  &  16,16 &  60,49 & R. Clyde & R & -4,31 &  55,56 & 5 & -8 & Al4 \\
5 & Lemannonius sinus & $\Omega$,$\Omega$  &  15,42 &  61,43 & Loch Fyne & T & -4,56 &  56,16 & 4 & 12 & Al4 \\
6 & Epidium pr. & $\Omega$,$\Omega$  &  14,16 &  61,15 & Mull of Kintyre & R & -5,45 &  55,17 & \text{---} & \text{---} & --- \\
7 & Longus fl.  & $\Omega$,$\Omega$  &  14,18 &  61,45 & Loch Linnhe & R & -5,38 &  56,29 & -16 & 0 & Al4 \\
8 & Itys fl.  & $\Omega$,$\Omega$  &  14,19 &  63,13 & Loch Alsh & T & -5,36 &  57,16 & 7 & -8 & Al1 \\
9 & Volsas sinus & $\Omega$,$\Omega$  &  14,37 &  64,12 & Loch Broom & T & -5,08 &  57,53 & -7 & -1 & Al1 \\
10 & Nabarus fl.  & $\Omega$,$\Omega$  &  14,34 &  64,41 & R. Naver & R & -4,14 &  58,32 & -7 & -7 & Al2 \\
11 & Tarvedum sive Orcas pr. & $\Omega$,$\Omega$  &  15,03 &  65,22 & Cape Wrath & B & -5,00 &  58,38 & 4 & 6 & Al1 \\
12 & Novantarum chersonesus & $\Omega$,$\Omega$  &  12,08 &  60,20 & Rhinns of Galloway & R & -5,10 &  55,00 & \text{---} & \text{---} & --- \\
13 & Abravannus fl.  & $\Omega$,$\Omega$  &  13,21 &  59,29 & Water of Luce & R & -4,49 &  54,52 & -34 & -2 & Al6 \\
14 & Iena ae. & $\Omega$,X  &  14,38 &  59,16 & R. Cree & R & -4,24 &  54,54 & -2 & -14 & Al6 \\
15 & Deva fl.  & $\Omega$,$\Omega$  &  15,09 &  58,44 & R. Dee & R & -4,04 &  54,50 & -25 & 7 & Al7 \\
16 & Novius fl.  & $\Omega$,$\Omega$  &  16,09 &  58,50 & R. Nith & R & -3,35 &  55,00 & -10 & 1 & Al7 \\
17 & Itunae ae. & $\Omega$,$\Omega$  &  17,37 &  58,48 & R. Eden & R & -3,04 &  54,57 & 24 & 3 & Al7 \\
42 & Virvedrum pr. & $\Omega$,$\Omega$  &  15,40 &  65,13 & Dunnet Head & --- & -3,22 &  58,40 & -10 & 8 & Al2 \\
43 & Verubium pr. & $\Omega$,$\Omega$  &  16,30 &  64,59 & Duncansby Head & B & -3,01 &  58,39 & 6 & -1 & Al2 \\
44 & Ila fl.  & $\Omega$,$\Omega$  &  16,31 &  64,44 & R. Wick & --- & -3,05 &  58,26 & 10 & 0 & Al2 \\
45 & Ripa alta & $\Omega$,$\Omega$  &  16,32 &  64,13 & Tarbat Ness & R & -3,47 &  57,52 & -3 & 0 & Al1 \\
46 & Loxa fl.  & X,X  &  17,17 &  63,28 & R. Lossie & R & -3,17 &  57,43 & -9 & 4 & Al3 \\
47 & Varar ae. & X,X  &  16,32 &  63,13 & Beauly Firth & R & -4,14 &  57,30 & 14 & 6 & Al3 \\
48 & Tuesis ae. & $\Omega$,$\Omega$  &  18,01 &  63,12 & R. Spey & R & -3,06 &  57,40 & 12 & -5 & Al3 \\
49 & Celnius fl.  & $\Omega$,$\Omega$  &  18,34 &  63,11 & R. Deveron & R & -2,31 &  57,40 & 2 & -5 & Al3 \\
50 & Taezalorum pr. & $\Omega$,$\Omega$  &  19,09 &  63,26 & Kinnairds Head & R & -2,00 &  57,42 & -3 & 4 & Al3 \\
51 & Deva fl.  & $\Omega$,$\Omega$  &  19,03 &  62,39 & R. Dee (at Aberdeen) & R & -2,05 &  57,08 & -3 & 2 & Al3 \\
52 & Tava ae. & $\Omega$,$\Omega$  &  18,57 &  62,08 & R. South Esk & T & -2,27 &  56,42 & 0 & 4 & Al4 \\
53 & Tina fl.  & $\Omega$,$\Omega$  &  18,51 &  61,36 & Firth of Tay & T & -2,47 &  56,27 & 16 & -4 & Al4 \\
54 & Boderia ae. & $\Omega$,X  &  17,39 &  60,53 & Firth of Forth & R & -2,39 &  56,08 & 3 & -5 & Al5 \\
55 & Alaunus fl.  & $\Omega$,$\Omega$  &  18,34 &  60,24 & R. Tweed & T & -1,59 &  55,46 & 3 & -5 & Al5 \\
66 & Lucopibia & $\Omega$,$\Omega$  &  14,38 &  59,16 & Wigtown & T & -4,27 &  54,52 & 1 & -12 & Al6 \\
67 & Rerigonium & $\Omega$,$\Omega$  &  14,05 &  59,52 & Stranrear & R & -5,01 &  54,54 & 10 & 13 & Al6 \\
68 & Carbantorigum & $\Omega$,$\Omega$ &  16,34 &  59,08 & --- & --- & \text{---} & \text{---} & \text{---} & \text{---} & --- \\
69 & Uxellum & $\Omega$,$\Omega$  &  16,30 &  58,53 & Roman fort at Ward Law & R & -3,32 &  54,59 & 2 & 5 & Al7 \\
70 & Corda & $\Omega$,$\Omega$  &  16,02 &  59,41 & Roman fort at Castledykes & R & -3,43 &  55,41 & -7 & -1 & Al7 \\
71 & Trimontium & $\Omega$,$\Omega$  &  17,12 &  59,06 & Newstead & R & -2,42 &  55,36 & -16 & -23 & Al7 \\
72 & Colania & $\Omega$,$\Omega$ &  17,05 &  59,53 & --- & --- & \text{---} & \text{---} & \text{---} & \text{---} & --- \\
73 & Vandogara & M,$\Omega$  &  15,31 &  60,23 & Ayr & T & -4,37 &  55,28 & -23 & 1 & Al4 \\
74 & Coria & $\Omega$,$\Omega$ &  16,52 &  60,24 & --- & --- & \text{---} & \text{---} & \text{---} & \text{---} & --- \\
75 & Alauna & $\Omega$,$\Omega$  &  17,00 &  61,02 & Stirling & T & -3,56 &  56,07 & 2 & -9 & Al4 \\
76 & Lindum & X,$\Omega$  &  16,41 &  61,11 & Drumquhassle & R & -4,27 &  56,04 & 19 & 0 & Al4 \\
77 & Victoria & $\Omega$,$\Omega$  &  17,46 &  61,24 & Kinross & H & -3,25 &  56,12 & 5 & 2 & Al4 \\
78 & Curia & X,$\Omega$  &  17,25 &  59,51 & Borthwick Castle & M & -3,00 &  55,50 & 12 & -2 & Al7 \\
79 & Alauna & X,X &  18,19 &  60,46 & --- & --- & \text{---} & \text{---} & \text{---} & \text{---} & --- \\
80 & Bremenium & $\Omega$,$\Omega$  &  17,59 &  60,05 & High Rochester & R & -2,16 &  55,17 & -6 & 10 & Al5 \\
81 & Banatia & $\Omega$,$\Omega$  &  16,45 &  61,41 & Roman fort at Dalginross & R & -3,59 &  56,22 & -6 & 5 & Al4 \\
82 & Tamia & $\Omega$,$\Omega$ &  17,10 &  62,11 & --- & --- & \text{---} & \text{---} & \text{---} & \text{---} & --- \\
83 & Pinnata castra & $\Omega$,$\Omega$  &  17,17 &  63,20 & Burghead & L & -3,29 &  57,42 & 3 & -1 & Al3 \\
84 & Tuesis & $\Omega$,$\Omega$  &  17,38 &  63,04 & Roman camp at Bellie & R & -3,06 &  57,37 & -5 & -7 & Al3 \\
85 & Orrea & $\Omega$,$\Omega$  &  18,20 &  61,38 & near Monifieth & R & -2,49 &  56,29 & -6 & -5 & Al4 \\
86 & Devana & $\Omega$,$\Omega$  &  18,31 &  62,48 & Roman camp at Kintore & R & -2,21 &  57,14 & -10 & 3 & Al3 \\
125 & Scitis insula & $\Omega$,$\Omega$  &  13,42 &  65,58 & Skye & R & -6,18 &  57,42 & \text{---} & \text{---} & --- \\
126 & Dumna insula & $\Omega$,X  &  12,37 &  64,38 & Lewis & R & -6,45 &  58,07 & 0 & 4 & Al1 \\
127 & Orcades insulae & $\Omega$,$\Omega$  &  11,51 &  64,36 & Orkney & R & -2,59 &  59,00 & \text{---} & \text{---} & --- \\
\hline
\end{longtable}
}

\begin{table} [h]
\caption{Results of the adjustment of the rotation of Ptolemaic Scotland}\label{tab:rotpar}
\centering
\begin{tabular}
{ >{\footnotesize}l >{\footnotesize}r >{\footnotesize}r >{\footnotesize}r >{\footnotesize}r >{\footnotesize}r >{\footnotesize}r}
\hline
$\lambda$ \& $\phi$ & \multicolumn{2}{>{\footnotesize}c}{1) Rivet \& Smith} & \multicolumn{2}{>{\footnotesize}c}{2) Strang} & \multicolumn{1}{>{\footnotesize}c}{3) Table \ref{tab:ortsliste}} \\
$\Lambda$ \& $\Phi$ & \multicolumn{1}{>{\footnotesize}c}{a) $\Omega$} &  \multicolumn{1}{>{\footnotesize}c}{b) X} & \multicolumn{1}{>{\footnotesize}c}{a) $\Omega$} &  \multicolumn{1}{>{\footnotesize}c}{b) X} & \\
\hline
 $\sigma_{\Lambda^*}$/$\sigma_{\Phi^*}$ & 1.2\degree/0.6\degree & 1.0\degree/0.5\degree & 1.2\degree/0.6\degree & 1.0\degree/0.5\degree & 0.8\degree/0.4\degree \\
 $n$                  & 47               & 47               & 50               & 50               & 45 \\
 $\Lambda_\mathrm{P}$ & $17\degree45' \pm  0\degree25'$ & $17\degree47' \pm  0\degree26'$ & $17\degree55' \pm  0\degree25'$ & $17\degree58' \pm  0\degree23'$ & $17\degree54' \pm  0\degree07'$ \\
 $\Phi_\mathrm{P}$    & $58\degree36' \pm  0\degree14'$ & $58\degree34' \pm  0\degree15'$ & $58\degree31' \pm  0\degree15'$ & $58\degree36' \pm  0\degree13'$ & $58\degree35' \pm  0\degree04'$ \\
 $\alpha$             & $-79\degree53' \pm  2\degree56'$ & $-80\degree34' \pm  2\degree38'$ & $-81\degree23' \pm  2\degree57'$ & $-82\degree48' \pm  2\degree24'$ & (const.) $-80\degree$ \\
 $\Lambda_0$          & $21\degree16' \pm  0\degree30'$ & $21\degree16' \pm  0\degree32'$ & $21\degree16' \pm  0\degree31'$ & $21\degree16' \pm  0\degree30'$ & $21\degree35' \pm  0\degree05'$ \\
 $\Phi_0$             & $-14\degree28' \pm  0\degree15'$& $-14\degree28' \pm  0\degree16'$ & $-14\degree27' \pm 0\degree15'$ & $-14\degree27' \pm  0\degree15'$ & $-14\degree28' \pm  0\degree04'$\\
 $s_0$                & 0.99             & 1.07             & 1.03             & 1.00             & 1.03 \\
\hline
\end{tabular}
\end{table}

\begin{table} [t]
\caption{Relative shifts of the transformation units (TU) in Scotland with respect to Al7 and scale-corrected a posteriori standard deviations}\label{tab:sch_trdiff}
\centering
\begin{tabular}
{ l r r r r @{ } r r @{ } r }
\hline
\multicolumn{1}{c}{TU} & \multicolumn{1}{c}{$n$} & \multicolumn{1}{c}{$\Delta \Lambda_{0k}$} & \multicolumn{1}{c}{$\Delta \Phi_{0k}$} & \multicolumn{2}{c}{$s_{\lambda^*}$} & \multicolumn{2}{c}{ $s_{\phi^*}$}\\
\hline
Al1 &  5 &    0\degree31$'$ & 1\degree33$'$ &     6$'$ & ( 6\;km) &     5$'$ & ( 9\;km)\\
Al2 &  4 & $-$0\degree46$'$ & 1\degree17$'$ &    10$'$ & (10\;km) &     6$'$ & (11\;km)\\
Al3 &  9 &    0\degree43$'$ & 0\degree54$'$ &     9$'$ &  (9\;km) &     5$'$ & ( 9\;km)\\
Al4 & 11 &    1\degree03$'$ & 0\degree56$'$ &    12$'$ & (12\;km) &     6$'$ & (11\;km)\\
Al5 &  3 & $-$0\degree02$'$ & 0\degree41$'$ &     5$'$ & ( 5\;km) &     9$'$ & (17\;km)\\
Al6 &  6 & $-$0\degree35$'$ & 0\degree55$'$ &    13$'$ & (14\;km) &    11$'$ & (20\;km)\\
Al7 &  7 & ---              & ---           &    15$'$ & (16\;km) &     8$'$ & (15\;km)\\
\hline
\end{tabular}
\end{table}

\clearpage

\begin{figure}[p]
\centering
\includegraphics[width=7cm]{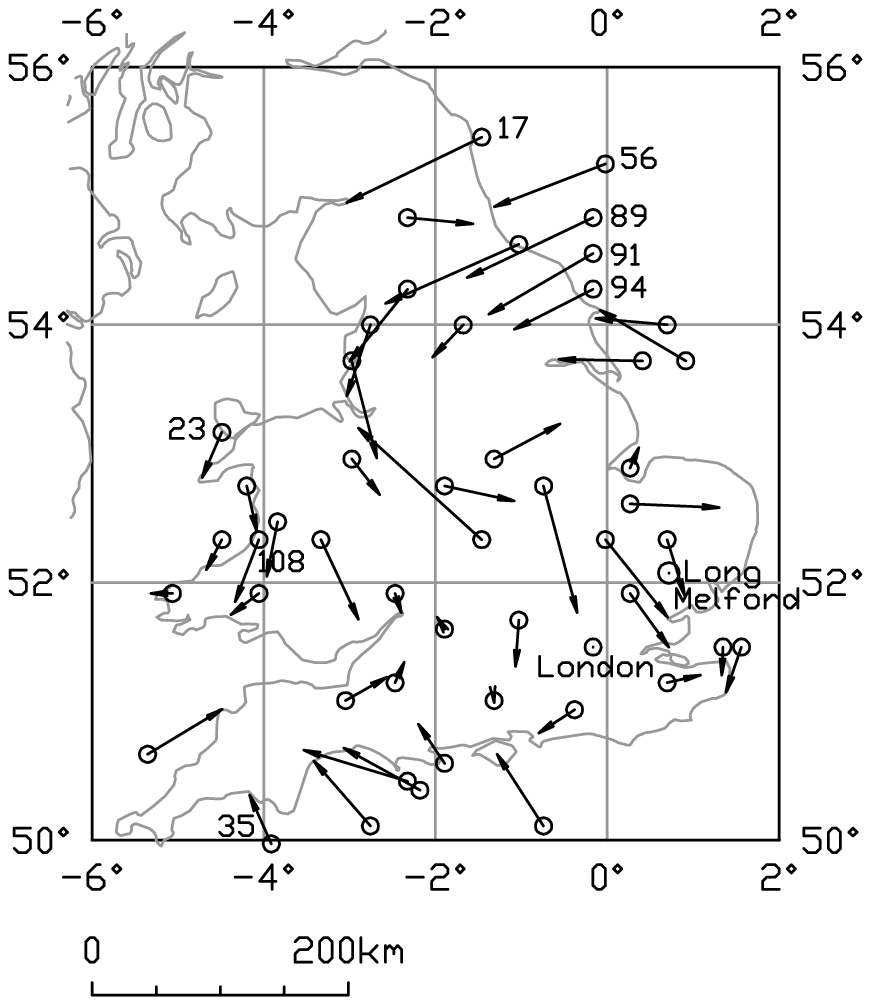}
\caption{Residual vectors between the transformed Ptolemaic positions (\textit{circle}) and the actual positions (\textit{arrowhead}) in England; the transformation is based on formula (\ref{eqn:trafo}) with $m_\Lambda=0.86436$, $m_\Phi=0.83333$ and a centring at London} \label{fig:strang}
\end{figure}

\begin{figure}[p]
\centering
\includegraphics[width=14cm]{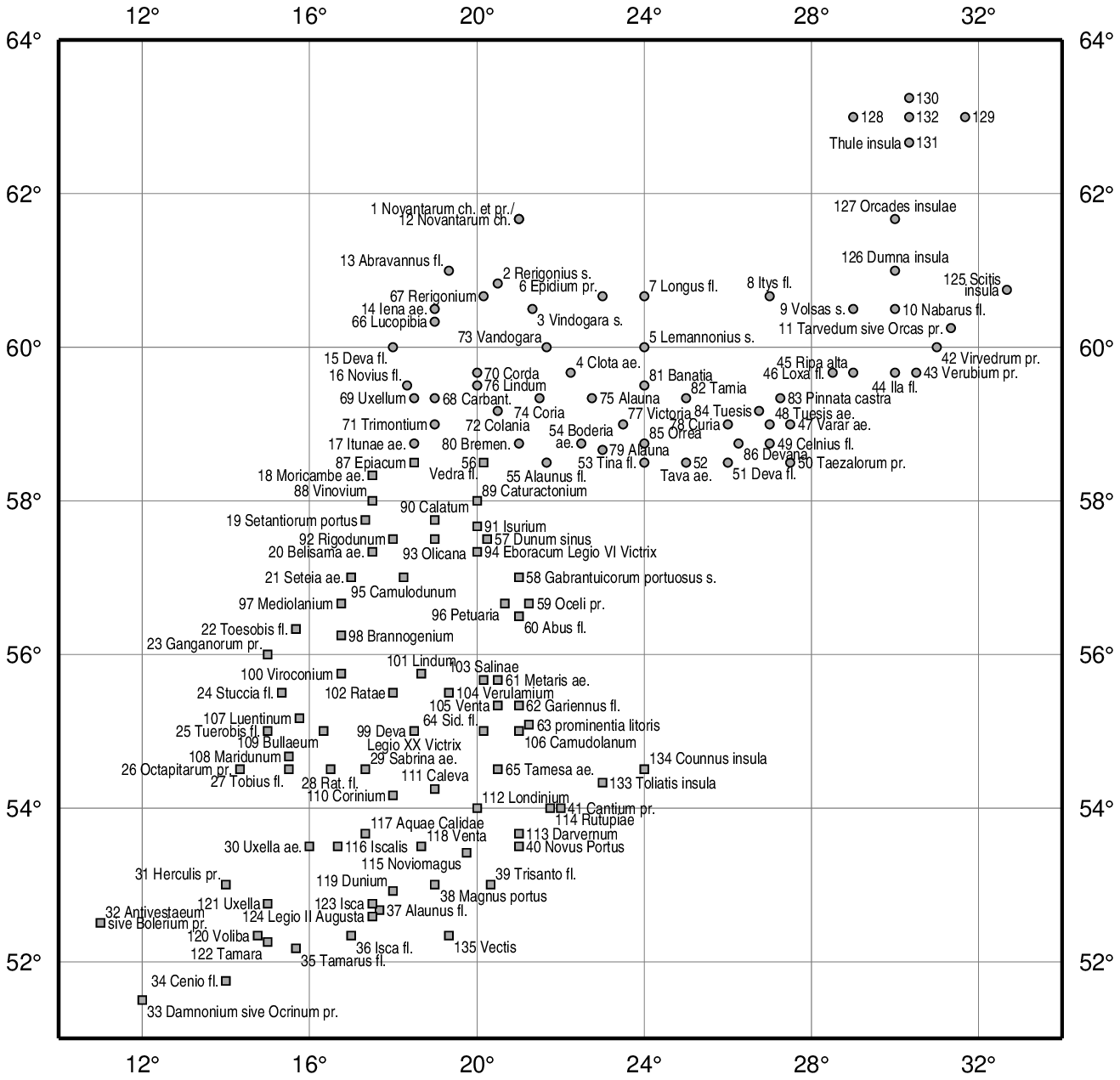}
\caption{Places of Ptolemaic England (\textit{square}) and Scotland (\textit{circle}) based on the $\Omega$-coordinates} \label{fig:albion}
\end{figure}

\begin{figure}[p]
\centering
\includegraphics[width=7cm]{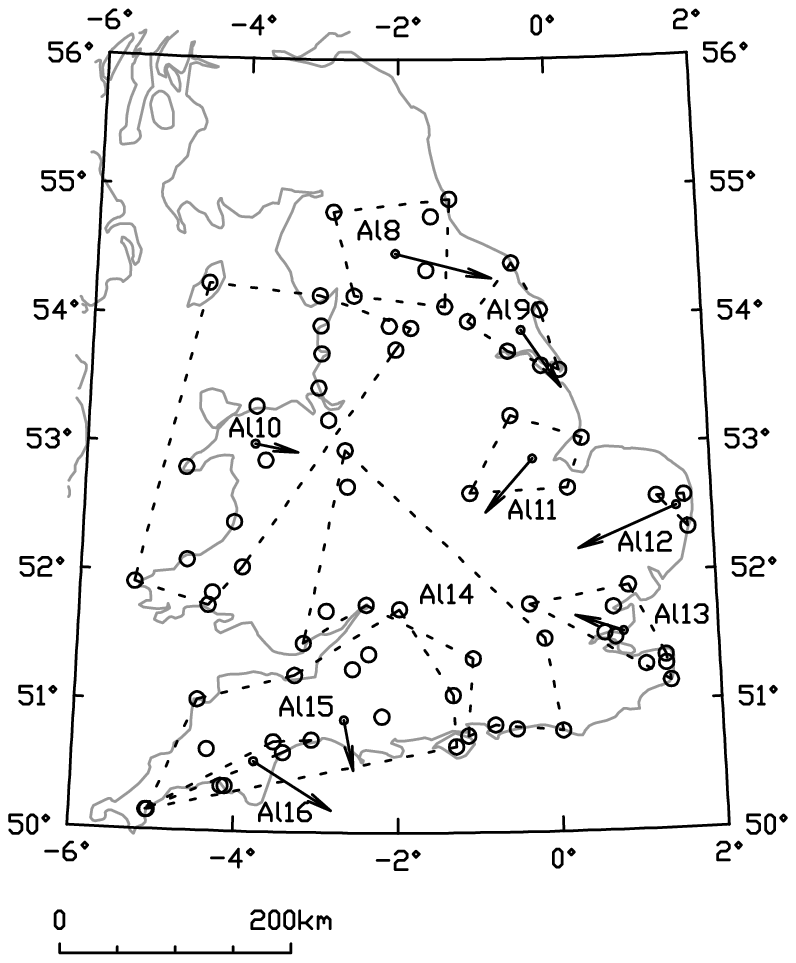}
\caption{Transformation units in England and their relative shifts with respect to Al14} \label{fig:eng_trdiff}
\end{figure}

\begin{figure}[p]
\centering
\includegraphics[width=7cm]{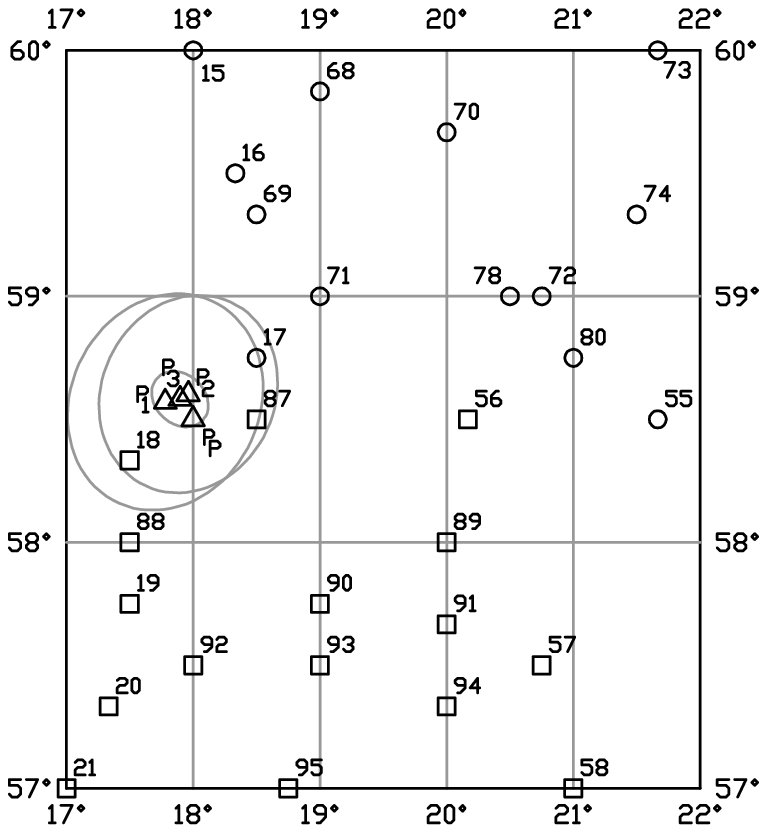}
\caption{Assumed pivot point P$_\mathrm{P}$, estimated pivot points P$_1$, P$_2$, P$_3$ (\textit{triangle}), their confidence ellipses, places of Ptolemaic England (\textit{square}) and Scotland (\textit{circle}) based on the X-coordinates (differences to $\Omega$ in Nos. 19, 20, 57, 68, 72, 78, 95)} \label{fig:ell}
\end{figure}

\begin{figure}[p]
\centering
\includegraphics[width=7cm]{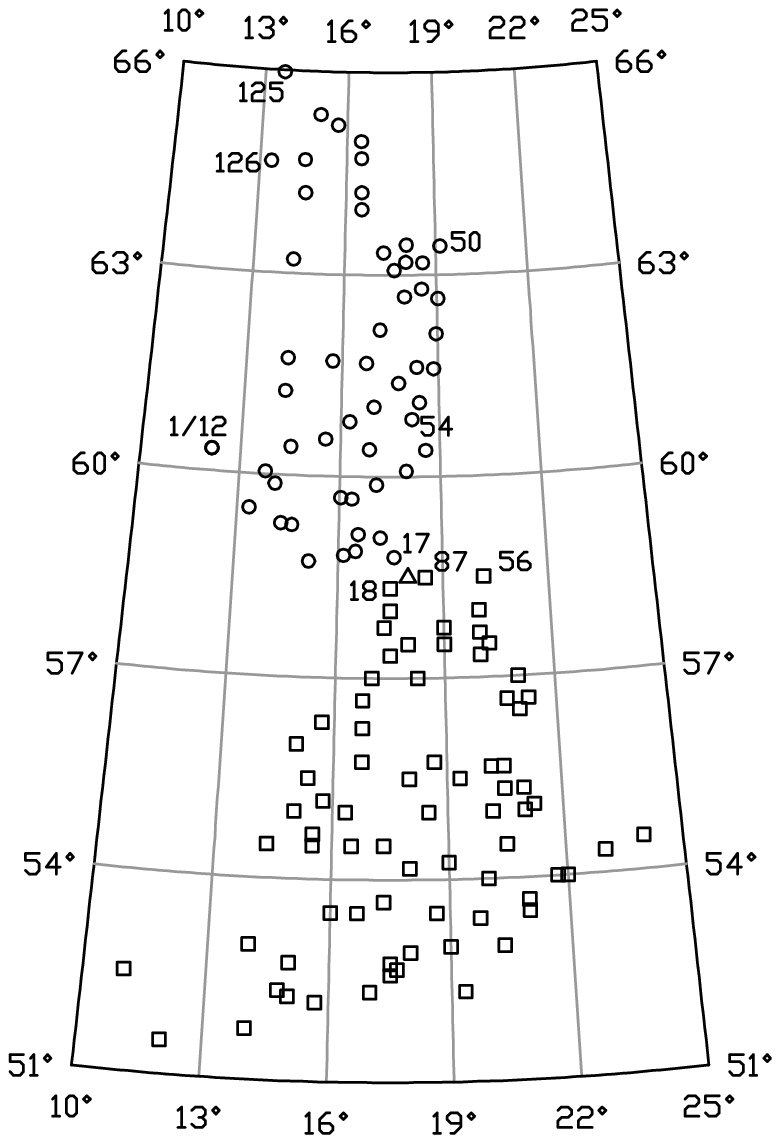}
\caption{Places of Ptolemaic England (\textit{square}) and Scotland (\textit{circle}) based on the $\Omega$-coordinates; the Scottish places are rotation-corrected by a rotation around P$_\mathrm{P}=(\Lambda=18\degree, \Phi=58\degree30')$ (\textit{triangle}) by $-\alpha_\mathrm{P}=80\degree$} \label{fig:bri_rot}
\end{figure}

\begin{figure} [p]
\centering
\subfigure[]{
\label{fig:coast_a}
\includegraphics[width=7cm]{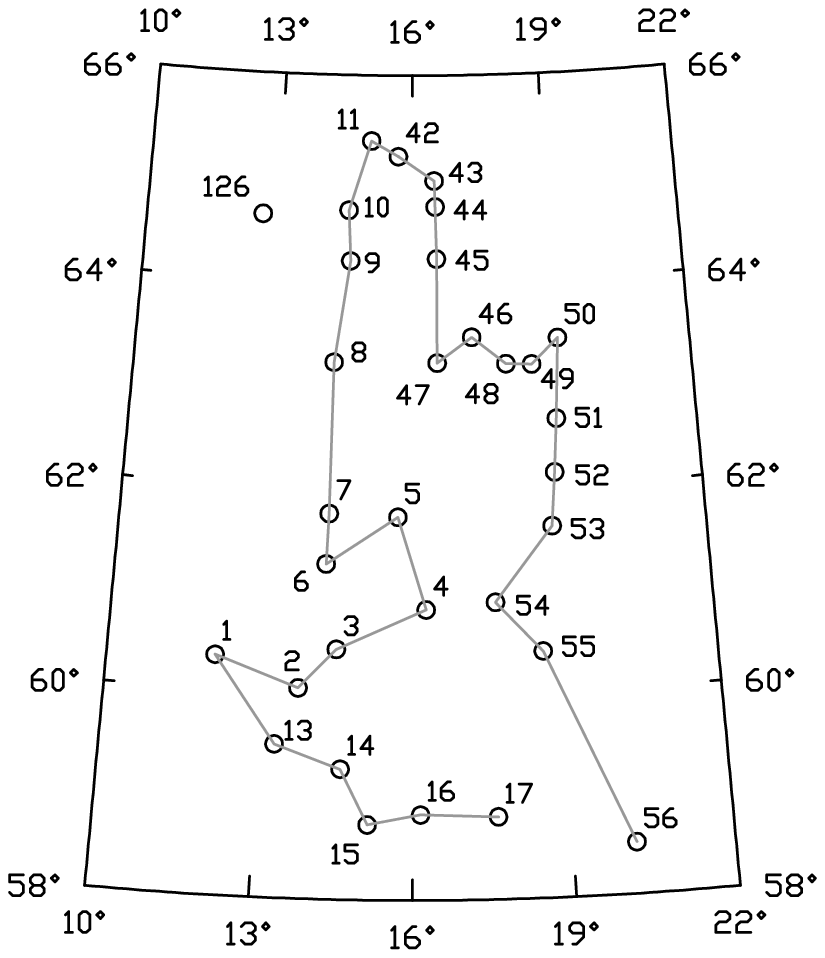}
}
\\
\subfigure[]{
\label{fig:coast_m}
\includegraphics[width=7cm]{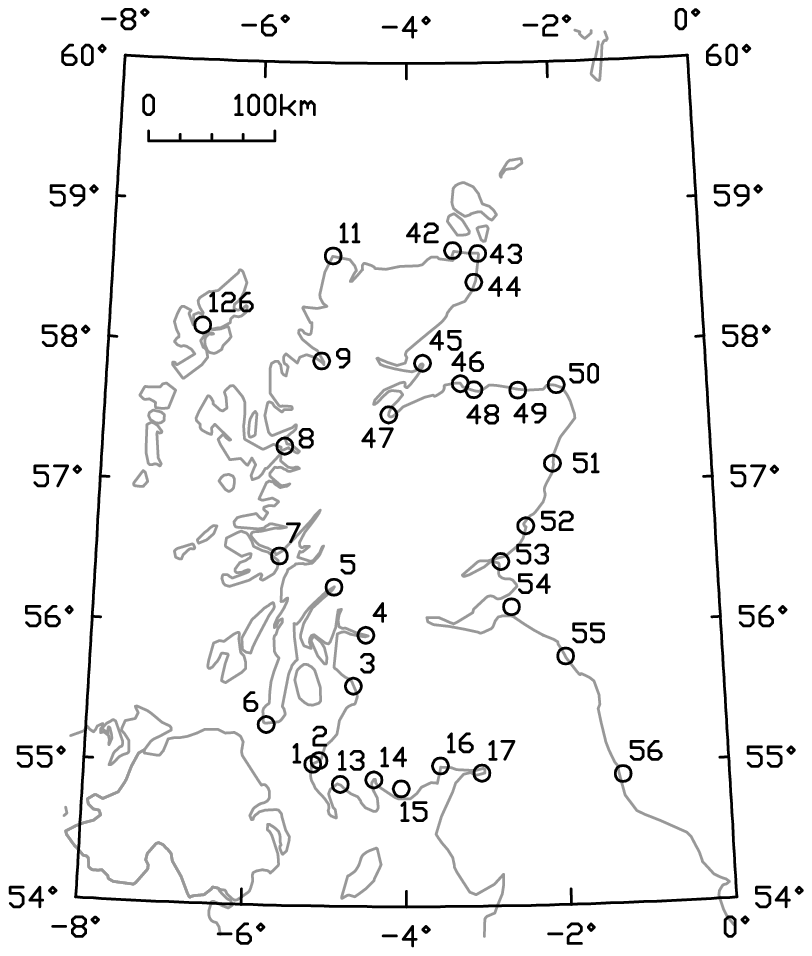}
}
\caption{Points and waters at the coast of Ptolemaic Scotland, (a) rotation-corrected ancient positions (cf. Table \ref{tab:ortsliste}), (b) modern positions and coast line (No. 56 belongs to Ptolemaic England)}
\end{figure}

\begin{figure}[p]
\centering
\includegraphics[width=7cm]{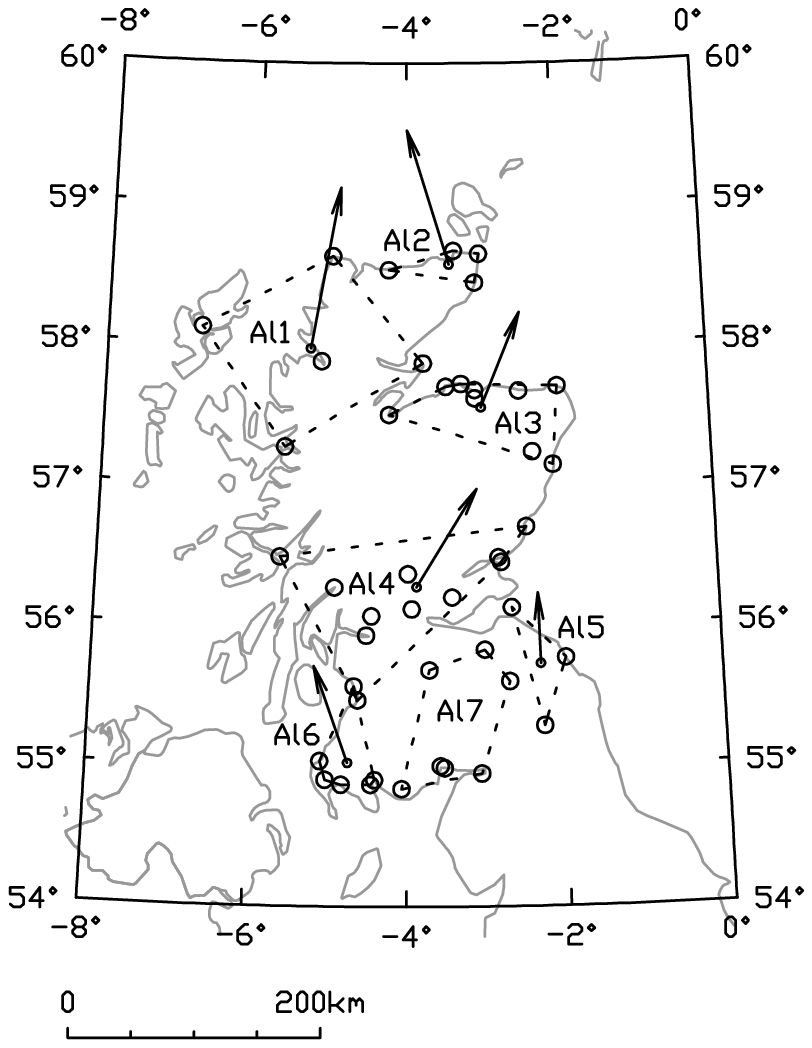}
\caption{Transformation units in Scotland and their relative shifts with respect to Al7} \label{fig:sch_trdiff}
\end{figure}

\begin{figure}[p]
\centering
\includegraphics[width=12.5cm]{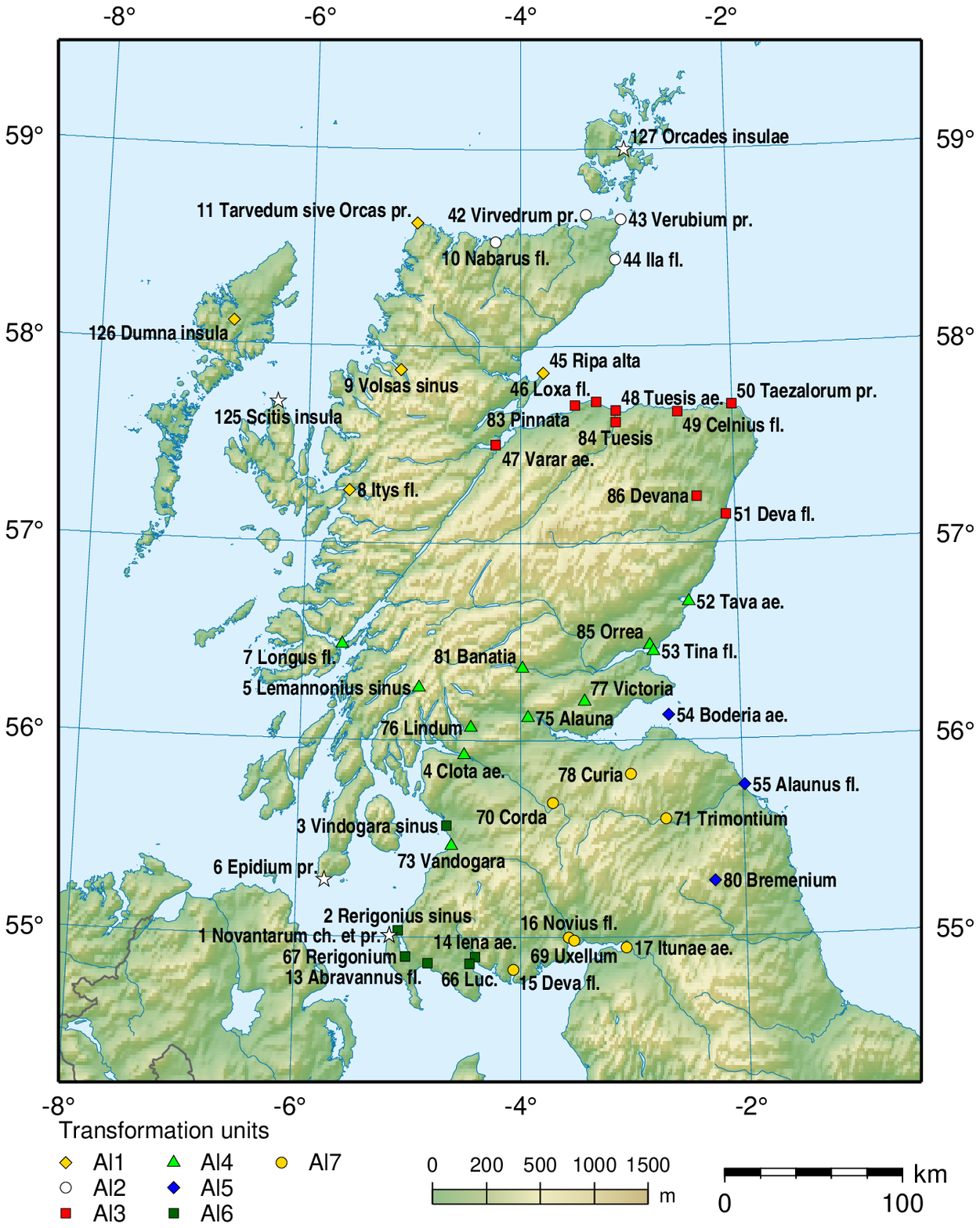}
\caption{Map of Ptolemaic Scotland (\textit{star}: no transformation unit)} \label{fig:karte}
\end{figure}

\end{document}